\documentstyle[amssymb,aps,epsf]{revtex}

\begin{document}
\draft
\title{Metric structure of random networks}
\author{S.N. Dorogovtsev$^{1, 2, \ast }$, J.F.F. Mendes$^{1,\dagger}$, and A.N.
Samukhin$^{1, 2, \ddag }$}
\address{$^{1}$ Departamento de F\'\i sica and Centro de F\'\i sica do Porto,\\
Faculdade de Ci\^{e}ncias, Universidade do Porto\\
Rua do Campo Alegre 687, 4169-007 Porto, Portugal\\
$^{2}$ A.F. Ioffe Physico-Technical Institute, 194021 St. Petersburg,\\
Russia}
\maketitle

\begin{abstract}
We propose a consistent approach to the statistics of the shortest paths in random graphs with a given degree distribution. This approach goes further than a usual tree ansatz and rigorously accounts for loops in a network. 
We calculate the distribution of shortest-path lengths (intervertex distances) in these networks and a number of related characteristics for the networks with various degree distributions. We show that in the large network limit this extremely narrow intervertex distance distribution has a finite width while the mean intervertex distance grows with the size of a network.  
The size dependence of the mean intervertex distance is discussed in various situations\vspace{9pt}. 
\end{abstract}


\section{Introduction}

The issue of discrete random geometries arises in numerous problems of quantum gravity \cite{bbpt97,bbj97,ckr98}, string theory, condensed matter physics (e.g., branched polymers \cite{adj90,aj95}), and classical statistical mechanics. In these problems, a fundamental question about the global structure of a random network (that is, a statistical ensemble of matrices) and its consequences naturally arises. This question exists even when the notion ``network'' is not used directly in the description of 
such a problem. As a simple example, we mention the backgammon (``balls-in-boxes'') model which was considered as a mean-field description of simplicial gravity \cite{bbj97}. Formally speaking, the formulation of this simple model does not contain the notion ``network''.  
However, the statistical ensembles that are produced by the backgammon model can be easily related to random networks with a complex distribution of 
connections, which is another representation of the model.  

In this paper, we study the global structural organization of a wide class of complex random networks, or speaking more strictly, we study their metric structure. 
An intervertex distance in a network is naturally defined as the length of the shortest path between a pair of vertices. So, the statistics of intervertex distances, that is an intervertex distance distribution, actually determine the metric structure of a random network. This distribution is the basic structural characteristic of random networks which are under extensive study by physicists for the last years (e.g. see Refs. \cite{s01,ab01a,dm01c,dmbook02}). Networks with fat-tailed degree distributions show a number of exciting effects (see Refs. \cite{ajb00d,cebh00,pv01}) and are especially intriguing. (By definition, degree is the total number of connections of a vertex, which is called sometimes ``the connectivity of a vertex''; a degree distribution is the distribution of degrees of vertices.) 

The intervertex distance distribution was obtained only for several very specific graphs. Even for basic uncorrelated random networks with a given degree distribution, the first moment of the intervertex distance distribution, that is the mean intervertex distance or the mean shortest-path length of a network, was only estimated \cite{nsw00,n02}. This estimation used the important fact that these networks are tree-like locally. This is not true at a large scale. 
In the recent paper \cite{ch02} the presence of loops was taken into account (see also Ref. \cite{pl01}) for estimating the mean intervertex distance of networks with a fat-tailed degree distribution.  

In this paper we propose a rigorous approach which takes into account both the locally tree-like structure of uncorrelated networks and the presence of loops on a large scale. This approach allows us to explicitly calculate the intervertex distance distribution and its moments, and to describe their dependence on the size of a network. 

Our approach is valid for uncorrelated random networks with
a given degree distribution. 
These basic ``equilibrium'' networks are graphs, which are maximally random  
under the constraint that their degree distribution is given. 
In graph theory these networks (loosely speaking, one of their versions) are called labelled random graphs with a given degree sequence or the configuration model \cite{bbk72,bc78,b80,w81}. These networks are a starting point for the study of the effects of complex degree distributions, and so are of fundamental importance. 

The (uncorrelated) random graph with a given degree distribution can be constructed in the following way. 
Take $N$ vertices. Attach to the vertices ``spines'', $\{q_i\}$, $i=1,\ldots,N$, according to a given sequence $\{N(q)\}$, $q=0,1,2\dots $, where $N=\sum_qN\left( q\right)$, so that the vertices look like a family of ``hedgehogs''. Connect various spines at random. 

This procedure provides the maximally
random graph with a given degree distribution $\Pi \left( q\right) =N\left(
q\right) /N$. Without lack of generality we can set the number of
zero-degree (i.e. isolated) vertices to be zero, $\Pi \left( 0\right) =0$. The main global topological properties of such a
networks are governed by the parameter \cite{nsw00,mr95,mr98}: 
\begin{equation}
z_1=\frac{\overline{q^2}}{\bar{q}}-1  
\, , 
\label{i10}
\end{equation} 
which is the ratio of the mean numbers of the second- and first-nearest neighbours of a vertex in the network. $\bar{q}$ and $\overline{q^2}$ are the first and the second moments of the degree distribution.  
For $z_1<1$, all the connected components of the network remain finite in the
infinite network limit (by definition, this is a thermodynamic limit). If $z_1>1$, the {\em giant connected
component} arises, whose size is proportional to the size of the whole network.
The condition for the emergence of the giant connected component \cite{mr95,mr98}, $z_1>1$, may be written as: 
\begin{equation}
\overline{q^2}-2\bar{q}=\sum_{q=1}^\infty q\left( q-2\right) \Pi \left(
q\right) =\sum_{q=2}^\infty q\left( q-2\right) \Pi \left( q\right) -\Pi
\left( 1\right) >0
\,.  
\label{i20}
\end{equation}
So, the giant connected component is formed only if the
fraction of ``dead ends'', $\Pi(1)$, is
sufficiently small. If ``dead ends'' are absent, giant connected
component exists and, in the thermodynamic limit, includes almost all 
vertices. (We do not consider the case when the network consists solely of
the vertices of degree two).

\begin{figure}
\epsfxsize=60mm
\centerline{\epsffile{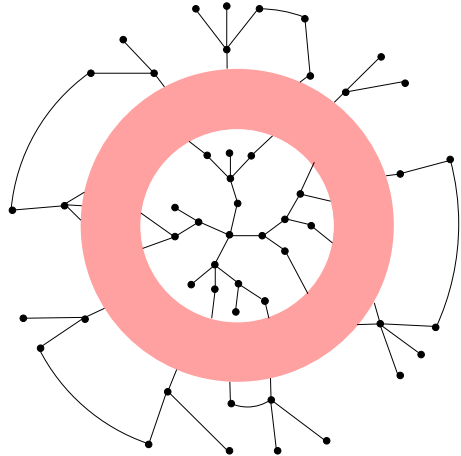}}
\caption{
Connected components of a vertex. Three first components, shown inside the
shaded area, are trees. The higher ones, shown outside the shaded area, are
assumed to contain a finite fraction of the network, and, therefore, may
contain closed loops.
}
\label{comp}
\end{figure}

We present a consistent approach allowing rigorous calculations of the
intervertex distance statistics within a giant connected component in such random networks. The main object we will consider is the sizes  
of connected components of a vertex in the graph (see a schematic view of the structure of an uncorrelated network in fig. \ref{comp}). The $n$-th order connected component of a vertex consists of all the vertices within the first $n$
coordination spheres of the vertex; in other words---the distance to
the central vertex in the $n$-th connected component does not exceed $n$.
Obviously, in a random graph the size of the connected component is a
fluctuating random variable. 


The idea of our method is to construct a
recurrent relation expressing the size distribution of the $n+1$-th
connected component through that of the $n$-th connected component. 
This relation can be derived in two limiting cases: when the size
of a connected component is negligibly small compared to the size of a network and when a connected component is a finite fraction of an infinitely large network.  
Sewing together
the results in these limiting cases yields the complete set
of connected component size distributions. In particular, this allows us to obtain the intervertex distance distribution in a random graphs.

The main results of this paper are as follows: 

\begin{itemize} 

\item[(1)] 
We find an explicit expression for the mean intervertex distance, $\ln N/\ln z_1+\mbox{const}$. In Ref. \cite{nsw00} this result was obtained as an estimate, here we present an exact result with an exact constant.  

\item[(2)] 
We obtain the form of the intervertex distance distribution and show that in the networks under
consideration, almost all vertices within a giant connected component
are nearly equidistant. More precisely, we found that the mean square
deviation of an intervertex distance is finite in the infinite network.  

\end{itemize}  


To find the intervertex distribution function, one has to solve the
functional equation, whose form is determined by the degree distribution
in the network. Sometimes this can be done explicitly (two examples are considered in the paper). 
However, even in a general case, 
all essential features of the distance distribution can be
reproduced analytically. 

First, the cumulative distance distribution $Q\left( d,N\right) $
(the probability that the intervertex distance is less than or equal to $d$),
appears to be actually the function of $l=d-\bar{d}\left( N\right) $, $%
Q\left( l,N\right) = {\cal Q}\left( l\right) $, where $\bar{d}\left( N\right)
=\ln \left( AN\right) /\ln z_1$ is the average intervertex distance, and $A$
is a number of the order of unity. 

Second, we find both the asymptotics
of ${\cal Q}\left( l\right) $ at large deviations $l$ of distances from the
mean value, positive and negative. At large negative $l$, this asymptotics
is determined by the first two moments of the degree distribution, $%
{\cal Q}\left( l\right) \sim z_1^l$. The asymptotics of ${\cal Q}\left(
l\right) $ on the other side---at large positive $l$---is determined by 
vertices with the lowest degrees. Obviously, ${\cal Q}\left( l\right) \rightarrow
m_\infty ^2$ as $l\rightarrow \infty $, where $m_\infty $ is the capacity of
the giant connected component, because $m_\infty ^2$ is precisely the probability
that a randomly chosen pair of vertices is interconnected. If the lowest degree
is either one or two, then the asymptotics of $m_\infty ^2-{\cal Q}\left(
l\right) $ at $l\rightarrow +\infty $ decreases exponentially with a linear
preexponential factor. If the lowest degree of the vertex is three or
higher, then the asymptotics decreases faster than an exponent. 

The scheme suggested herein works only if the parameter $z_1$ is finite, which means the
convergence of the first and second moments of the degree distribution in the thermodynamic limit. 
This is not the case if this degree distribution asymptotically behaves as $%
\Pi \left( q\right) \sim q^{-\gamma }$ with $\gamma  \leq3$ at large degrees $q$%
. We have studied the case of $\Pi \left( q\right) \sim q^{-\gamma }\exp
\left( -q/q_0\right) $, $2<\gamma <3$ with large but finite value of the
cut-off parameter $q_0$. As a result, we have found that in this case $\bar{d}%
\sim \ln N/q_0$, and ${\cal Q}\left( l\right) $ is independent both
of the system size and the cut-off parameter. Again, the mean square
deviation of intervertex distances is of the order of unity, and all the 
vertices in the giant connected component are nearly equidistant. 

This
result is valid in the limit $N\rightarrow \infty $, when we assume that the
cut-off parameter is large. In reality, in the finite-size networks
a degree distribution has some natural size-dependent cut-off. How the cut-off
parameter varies with the size of the network, depends on the details of a 
construction procedure. For example, in the configuration model the position of the cut-off depends
on how the limit $N\left( q\right) /N\rightarrow \Pi \left( q\right) $ is
approached. We show that the picture described above remains valid, if the
cut-off parameter grows with the network size sufficiently slowly, namely, not
faster than $\ln q_0 \sim \ln N/\ln \ln N$. Thus, for ``scale-free'' networks
with $2<\gamma <3$, we arrive at the stable distribution of
intervertex distances around their steadily growing mean value $\bar{d}(N)$ if
the construction procedure ensures not very fast growth of the degree
distribution cut-off with the network size.  
In this situation, $\bar{d}$ grows with $N$
slower than $\ln N$ but faster than $\ln \ln N$.

The paper is organized as follows. In Section \ref{defs} we define main
notions. In Section \ref{micr} we remind and essentially 
refine the approach of Ref. \cite{nsw00}, which is based on the tree ansatz and valid for finite-size
connected components in the infinite network. In Section \ref{macr}
we present the recursion relation between the sizes of the $n$-th and $n+1$-th
connected components in the limit, when these sizes are both infinite, taking into account loops. In
Section \ref{sawing} we explain, how the results of two previous sections can be sewed together in the region, where the size of a connected component
is large compared to unity but small compared to the size of the whole
network. In Section \ref{isum} the results of previous sections are briefly
summed up and general results for various quantities of interest are 
presented.  
In Section \ref{exp}, as an illustrative example, we present an exact analytical solution for
the uncorrelated network with the degree distribution $\Pi \left( q\right) =Cq^{-1}\zeta ^q$, $\zeta <1$%
. In Section \ref{power} the network with the degree distribution $\Pi \left( q\right) \sim q^{-\gamma }\zeta
^{q} $, $2<\gamma <3$ and $\zeta <1$, $1-\zeta \ll 1$ is studied. In Section \ref{concl} we summarize the results obtained in the paper and discuss the size dependence $\bar{d}(N)$ in situations when it grows slower than $\ln N$, e.g. 
as $\ln\ln N$ or $\ln N/ \ln\ln N$.  
Some technical details are presented in two Appendices.

\section{Definitions}
\label{defs}

A graph consists of vertices connected by edges. Undirected
graph is described by its symmetric adjacency matrix $\hat{a}$. Elements of
this matrix are either $a_{ij}=a_{ji}=1$, if vertices $i$ and $j$ are
connected, or $a_{ij}=a_{ji}=0$ otherwise. We consider only graphs with $%
a_{ii}=0$, that is ones without ``tadpoles''---edges with both ends
attached to the same vertex. The degree of a vertex, $q_i$, (sometimes it is
called the vertex connectivity) is the number of edges, attached to the vertex: $%
q_i=\sum_ja_{ij}$. Random networks are usually described in terms of a statistical 
ensemble: the set of graphs $G$ with corresponding statistical weights---a non-negative function $P\left( g\right) $, $g\in G$, defined on this set \cite{bck01,k01,dms02,bl02}.

Let us consider a statistical ensemble of undirected graphs, each of which contains $%
N\rightarrow \infty $ vertices. Let us choose an ensemble characterized by
a degree distribution $\Pi \left( q\right) $ and
maximally random otherwise. Several ensembles, equivalent in the
thermodynamic limit $N\rightarrow \infty $, may be used \cite{dms02}. For
example, one can use a ``microcanonical one'', usually referred to as the 
``configuration model''. 
Here we ascribe equal statistical weights to all
possible graphs with $N=\sum_qN\left( q\right) $ vertices, $N\left( q\right) 
$ of them have a degree $q$, $q=1,2,\dots $ (without lack of generality
we can exclude the possibility that a vertex is of zero degree). We assume,
that in the thermodynamic limit, $N\rightarrow \infty $, $N\left( q\right)
/N\rightarrow \Pi \left( q\right) $. For this ensemble, we have the degree distribution: 
\begin{equation}
\left\langle \delta _K\left( q_i-q\right) \right\rangle =\frac 1N%
\left\langle \sum_{i=1}^N\delta _K\left( q_i-q\right) \right\rangle =\Pi
\left( q\right) 
\,.  
\label{4}
\end{equation}
One can show that in the thermodynamic limit even degrees of the nearest-neighbour
vertices are uncorrelated in such networks: 
\begin{equation}
\frac 1{2L}\left\langle \sum_{i\ne j=1}^N\delta _K\left( q_i-q\right)
a_{ij}\delta _K\left( q_j-q^{\prime }\right) \right\rangle =\frac{qq^{\prime
}}{\bar{q}^2}\Pi \left( q\right) \Pi \left( q^{\prime }\right) 
\,,  
\label{6}
\end{equation}
where $\bar{q}=2L/N$ is the average vertex degree. We introduced the
notation $\delta _K\left( q-q^{\prime }\right) $ for the Kronecker symbol $\delta
_{qq^{\prime }}$. Relation (\ref{6}) plays a crucial role. In fact, the 
scheme presented here is based on this relation. 

We call the set of vertices, for which the shortest distance from some
vertex equals $n$, the $n$-th {\em shell} of this vertex. The union of the shells of a vertex 
from zeroth to $n$-th one inclusively is called the $n$-th (connected) {\em %
component} of the vertex.

Following \cite{nsw00}, it is convenient to use the degree distribution in
$Z$-representation (sometimes this object is called the generating function of the
distribution): 
\begin{equation}
\phi \left( x\right) =\frac 1N\sum_{i=1}^N\left\langle x^{q_i}\right\rangle
=\sum_{q=0}^\infty \Pi \left( q\right) x^q
\,.  
\label{10}
\end{equation}

Another useful quantity is what may be called an ``edge multiplication''
distribution function $\Pi _1\left( q\right) $. This is the conditional
probability that in a connected pair of vertices, a vertex has its
degree equal to $q+1$: 
\begin{equation}
\Pi _1\left( q\right) =\frac{\left\langle a_{ij}\delta _K\left(
q_i-q-1\right) \right\rangle }{\left\langle a_{ij}\right\rangle }=\frac{%
\left\langle \sum_ja_{ij}\delta _K\left( q_i-q-1\right) \right\rangle }{%
\left\langle \sum_ja_{ij}\right\rangle }=\frac{\left( q+1\right) \Pi \left(
q+1\right) }{\bar{q}}
\,,  
\label{15}
\end{equation}
or, in $Z$-representation,  
\begin{equation}
\phi _1\left( x\right) =\sum_q\Pi _1\left( q\right) x^q=\frac{\left\langle
a_{ij}x^{q_i-1}\right\rangle }{\left\langle a_{ij}\right\rangle }=\frac{%
\left\langle q_ix^{q_i-1}\right\rangle }{\bar{q}}=\frac{\phi ^{\prime
}\left( x\right) }{\phi ^{\prime }\left( 1\right) }
\,.  
\label{20}
\end{equation}
We make use that in our ensemble all pairs of vertices are statistically
equivalent. $\Pi _1\left( q\right) $ may also be thought of as the
probability that, choosing a random {\em edge} (but not a vertex!), and going along 
it in some of directions, we arrive at a vertex which has a degree 
equal to $q+1$ and therefore, there exist $q$ different possibilities to move
further.

\section{Microscopic components}
\label{micr}

By ``microscopic components'' of a vertex we mean the components of a size negligible compared to the size of the network.    

The role of $\phi _1\left( x\right) $ may be understood from the following
reasoning. Let us choose a random vertex $i$ of degree $q_i$. Assume that its $n$%
-th connected component is a tree. Then it consists of $q_i$ trees generated
by every edge attached to the vertex. Let $S_n^{\left( j\right) }$ $%
j=1\dots q_i$ be the number of vertices in such a tree. Obviously, the total
number of vertices in the $n$-th component $M_n=1+\sum_{j=1}^{q_i}S_n^{\left(
j\right) }$, and $S_n^{\left( j\right) }$, by the definition of
the statistical ensemble under consideration, are equally distributed, independent (in the
thermodynamic limit) random variables. For the distribution of $M_n$, we have
in $Z$-representation: 
$$
\Phi _n\left( x\right) =\left\langle
x^{M_n}\right\rangle =x\left\langle \left( x^{S_n}\right) ^q\right\rangle
=x\phi \left[ F_n\left( x\right) \right]
\, , 
$$ 
where $F_n\left( x\right)
=\left\langle x^{S_n}\right\rangle $ is the distribution function of $S_n$,
the number of vertices in the $n$-th order tree, formed by a randomly chosen
edge. We have: $S_{n+1}=1+\sum_{j=1}^{q_1}S_n^{\left( j\right) }$, where the
distribution function of $q_1$ is $\phi _1(x)$ in $Z$-representation. Then we
obtain finally for the size distribution of the $n$-th component: 
\begin{eqnarray}
\Phi _n\left( x\right) & = & x\phi \left[ F_n\left( x\right) \right]
\,, 
\nonumber
\\[5pt]
\;F_{n+1}\left( x\right) & = & x\phi _1\left[ F_n\left( x\right) \right]
\,,\ F_1\left( x\right) =x
\,.  
\label{30}
\end{eqnarray}

As $n\rightarrow \infty $ and $\left| x\right| <1$, $F_n\left( x\right)
\rightarrow F\left( x\right) $, where $F\left( x\right) $ describes the
size distribution of {\em finite} components attached to a randomly
chosen edge. Then $H\left( x\right) =\phi \left[ F\left( x\right) \right] $
is the finite-component size distribution. Note that $t_c=F\left(
1\right) \equiv \lim_{x\rightarrow 1-0}F\left( x\right) $ is the stable
fixed point of the recursion relation $t_{n+1}=\phi _1\left( t_n\right) $.
Taking into account that $\phi _1\left( x\right) $ is monotonously
increasing and convex downward as $0<x<1$, and $\phi _1\left( 1\right) =1$,
one can conclude that $t_c=1$ if $\phi _1^{\prime }\left( 1\right) \equiv
z_1\le 1$, and $t_c<1$ if $z_1>1$. In the latter case we have $H\left(
1\right) =\phi \left( t_c\right) <1$. But $H\left( 1\right) $ is the
probability that a randomly chosen vertex belongs to some finite component.
This means that as  
\begin{equation}
z_1=\phi _1^{\prime }\left( 1\right) =\frac{\phi ^{\prime \prime }\left(
1\right) }{\phi ^{\prime }\left( 1\right) }>1\,,  \label{35}
\end{equation}
a {\em giant connected component} appears in the network. Its capacity (the
probability that a vertex belongs to the giant component) is  
\begin{equation}
m_\infty =1-\phi \left( t_c\right) 
\,.  
\label{37}
\end{equation}

The average number of vertices in the $n$-th connected component of a vertex is $%
\left\langle S_n\right\rangle =\Phi _n^{\prime }\left( 1\right)
=1+z_0F_n^{\prime }\left( 1\right) $, $z_0\equiv \phi ^{\prime }\left(
1\right) $. One can easily find from Eq.~(\ref{30}) that $F_n^{\prime
}\left( 1\right) =\left( z_1^n-1\right) /\left( z_1-1\right) \sim z_1^n$ as $%
n\rightarrow \infty $, where $z_1=\phi _1^{\prime }\left( 1\right) =\phi
^{\prime \prime }\left( 1\right) /\phi ^{\prime }\left( 1\right) $. Let us
introduce, instead of $F_n$, the sequence of functions $f_n$ that is defined as  
\[
F_n\left( x\right) =f_n\left( \frac{z_1^n-1}{z_1-1}\ln \frac 1x\right) \,. 
\]
The recursion relation then turns into   
\[
f_{n+1}\left( y\right) =\exp \left( -\frac{z_1-1}{z_1^{n+1}-1}y\right) \phi
_1\left[ f_n\left( y\frac{z_1^n-1}{z_1^{n+1}-1}\right) \right]
\,,\;f_1\left( y\right) =e^{-y}. 
\]
As $n\rightarrow \infty $, it may be replaced with  
\begin{equation}
f_{n+1}\left( y\right) =\phi _1\left[ f_n\left( y/z_1\right) \right] 
\, , \ \ 
f_1(y)=e^{-y}  
\,.
\label{38}
\end{equation}
From $\phi _1\left( 1\right) =1$ it follows that $f_n\left( 0\right)
=F_n\left( 1\right) =1$. Also, we have now $f_n^{\prime }\left( 0\right) =-1$
independent of $n$. Taking into account that $\phi _1\left( x\right) $ is
analytic, monotonically growing and convex downward at $0<x<1$, one can
prove that the sequence $f_n$ converges as $n\rightarrow \infty $ to some
function $f(y)$. This latter may be found from the stationarity condition: 
\begin{equation}
f\left( y\right) =\phi _1\left[ f\left( y/z_1\right) \right] 
\,;\;
f\left(
0\right) =1\,,\;f^{\prime }\left( 0\right) =-1
\,.  
\label{40}
\end{equation}
The above conditions determine $f\left( y\right) $ uniquely. One can check
this, e.g. taking subsequent derivatives of Eq.~(\ref{40}) at $y=0$, which
allows us to express $f^{\left( k\right) }\left( 0\right) $ through $f^{\left(
l\right) }\left( 0\right) $, $l<k$. Then we have asymptotically at $%
n\rightarrow \infty $: 
\begin{equation}
\Phi _n\left( x\right) \rightarrow \phi \left[ f\left( \frac{z_1^n}{z_1-1}%
\ln \frac 1x\right) \right] 
\,.  
\label{45}
\end{equation}
The distribution function for the size of $n$-th connected component, $M_n$, in
the usual representation, $P_n\left( M\right) $, is the inverse $Z$-transform of 
$\Phi _n$: 
\begin{equation}
P_n\left( M\right) \equiv \oint_{\left| x\right| =1-\delta }\frac{dx}{2\pi i}%
\Phi _n\left( x\right) x^{-M-1}
\,.  
\label{47}
\end{equation}
Taking into account Eq.~(\ref{40}), we have in the limit $n\rightarrow
\infty $: 
\begin{eqnarray}
P_n\left( M\right) & \rightarrow & 
\left( z_1-1\right) z_1^{-n}p\left[ \left(
z_1-1\right) z_1^{-n}M\right] 
\,;
\nonumber
\\[5pt]
p\left( s\right) & = & \int_{-i\infty +\delta
}^{+i\infty +\delta }\frac{dy}{2\pi i}g\left( y\right) e^{sy}\,,\;g\left(
y\right) =\phi \left[ f\left( y\right) \right] 
\,.  
\label{50}
\end{eqnarray}
Note the order of limiting transitions adopted in this section: $%
\lim_{n\rightarrow \infty }\lim_{N\rightarrow \infty }$. The first is the 
thermodynamic limit, and only then the order of a connected component, $n$, is
tended to infinity.

Several essential assumptions had been made during the derivation of
the above formulae. First, it was required that $\phi ^{\prime }\left( 1\right) 
$ and $\phi ^{\prime \prime }\left( 1\right) $ are finite, which means that
the degree distribution $\Pi \left( q\right) $ has finite first and second
moments. This will be assumed everywhere below. And second, the graph must 
be a tree. In fact, it is sufficient that the $n${\em -th connected component}
of a vertex is a tree. But {\em almost all }$n{\em -th}${\em \ components
of infinite uncorrelated random graph are trees at finite }$n${\em .} Indeed, the
probability that two vertices in the $n$-th component are connected by an
edge is proportional to the ratio of the total numbers of edges inside and
outside this component. If $n$ is fixed and $N\rightarrow \infty $, this
ratio scales as 
$M_n/N$. Our final result for the component size
distribution, Eqs. (\ref{40}), (\ref{50}), is valid in the limit $n\rightarrow \infty $%
, which must be taken {\em after} the ``thermodynamic'' limit $N\rightarrow
\infty $. We emphasize that the order of limits is extremely essential here.

\section{Macroscopic components}
\label{macr}

Now we assume a different situation: the size of the graph $N$ and the order
of a connected component, $n$, simultaneously tend to infinity. At the same
time, we assume that the distribution of the capacity, $m_n=S_n/N$, of the $n$-th connected
component, $p_n\left( m\right) $, tends to some limiting $N$%
-independent distribution. From the results of the previous section it follows that
we have to assume that $z_1^n/N$ remains constant.

In this case the $n$-th connected component is no longer a tree. However, in
this case it appears to be possible to derive an exact (in the
thermodynamic limit) relation between $p_n\left( m\right) $ and $%
p_{n+1}\left( m\right) $. The idea is to use the law of large numbers.
Assume we have the $n$-th connected component with $M_n=Nm_n$ vertices. Also, we assume that the number of edges, which connect vertices inside the $n$-th
component to vertices outside this 
component is $L_n=Nl_n$. Due to the randomness of the graph, $m_{n+1}$
and $l_{n+1}$ would be fluctuating variables even if $m_n$ and $l_n$ are
fixed. However, in the thermodynamic limit fluctuations of intensive
variables $m_{n+1}$ and $l_{n+1}$ tend to zero. So, the evolution of the $n$-th
connected component, as $n$ is growing, is governed by 2-$d$ mapping: $\left(
m_n,l_n\right) \rightarrow \left( m_{n+1},l_{n+1}\right) $.

This mapping may be constructed as follows. Let $\Pi _n\left( q\right) $, or $%
\phi _n\left( y\right) $ in $Z$-representation, be the degree distribution of
vertices {\em outside} the $n$-th component. 
(Do not mix $\phi_{n=1}(y)$ and $\phi_1(y) \equiv \phi'(y)/\phi'(1)$.)   
Their total degree is $%
N\left( 1-m_n\right) \phi _n^{\prime }\left( 1\right) $. $Nl_n$ edges of
this number are to be chosen to connect with vertices of the $n+1$-th shell. All
such choices are equiprobable, because of the nature of the statistical
ensemble of graphs under consideration. So, the probability
that a vertex of a degree $q$  
outside the $n$-th component is not connected
to a vertex inside the $n$-th component equals $\left( 1-c_n\right) ^q$, where $%
c_n=l_n/\left[ \left( 1-m_n\right) \phi _n^{\prime }\left( 1\right) \right] $%
. Then the fraction of vertices, remaining outside the $n+1$-th shell, $1-m_{n+1}$%
, is given by  
\begin{equation}
\frac{1-m_{n+1}}{1-m_n}=\sum_{q=0}^\infty \Pi _n\left( q\right) \left(
1-c_n\right) ^q=\phi _n\left( 1-c_n\right) 
\,.  
\label{60}
\end{equation}
Also, we have a recursive relation for the degree distribution function: 
\begin{equation}
\Pi _{n+1}\left( q\right) =\frac{1-m_n}{1-m_{n+1}}\Pi _n\left( q\right)
\left( 1-c_n\right) ^q
\,,  
\label{70}
\end{equation}
or, in $Z$-representation,  
\begin{equation}
\phi _{n+1}\left( y\right) =\frac{\phi _n\left[ \left( 1-c_n\right) y\right] 
}{\phi _n\left( 1-c_n\right) }
\,.  
\label{80}
\end{equation}
Repeatedly applying Eq.~(\ref{80}), introducing $t_n=\left(
1-c_{n-1}\right) \left( 1-c_{n-2}\right) \cdots \left( 1-c_1\right)$, and 
using $\phi_{n=1}(x)=\phi(x) $, one
can write:  
\begin{equation}
\phi _n\left( y\right) =\frac{\phi \left( t_ny\right) }{\phi \left(
t_n\right) }
\,.  
\label{90}
\end{equation}
From Eqs.~(\ref{60}) and (\ref{90}) we obtain the relation  
\begin{equation}
1-m_n=\phi \left( t_n\right) 
\,,  
\label{100}
\end{equation}
relating $t_n$ and $m_n$. From the definition of $t_n$ we obtain the following
equation: 
\begin{equation}
t_{n+1}=\left( 1-c_n\right) t_n=t_n-\frac{t_nl_n}{\left( 1-m_n\right) \phi
_n^{\prime }\left( 1\right) }=t_n-\frac{l_n}{\phi ^{\prime }\left(
t_n\right) }
\,,  
\label{110}
\end{equation}
where Eqs.~(\ref{90}), (\ref{100}), and the definition of $c_n$ were used. Eqs.~(\ref{100}) and (\ref
{110}) express $m_{n+1}$ in terms of $m_n$ and $l_n$.

The total degree of vertices outside the $n$-th component may be written as $N\left(
1-m_n\right) \phi _n^{\prime }\left( 1\right) =Nt_n\phi ^{\prime }\left(
t_n\right) $ and outside the $n+1$-th one---as $Nt_{n+1}\phi ^{\prime }\left(
t_{n+1}\right) $. Therefore, the total degree of vertices in the $n+1$ shell is $%
N\left[ t_n\phi ^{\prime }\left( t_n\right) -t_{n+1}\phi ^{\prime }\left(
t_{n+1}\right) \right] $. Of this number, $Nl_n$ edges are attached to vertices
in the $n$-th component. Each of remaining ``free'' $N\left[ t_n\phi ^{\prime
}\left( t_n\right) -t_{n+1}\phi ^{\prime }\left( t_{n+1}\right) -l_n\right] $
edges may be attached either to a vertex outside the $n+1$-th component, or
to some other vertex in the $n+1$-th shell (see fig. \ref{comp}). 
The respective probabilities
relates as the total degree outside the $n+1$-th component, $Nt_{n+1}\phi ^{\prime
}\left( t_{n+1}\right) $, and the number of ``free'' edges in the $n+1$-th shell. So,
we have for the number of edges $Nl_{n+1}$, going out from the $n+1$-th component: 
\begin{equation}
l_{n+1}=\left[ t_n\phi ^{\prime }\left( t_n\right) -t_{n+1}\phi ^{\prime
}\left( t_{n+1}\right) -l_n\right] \frac{t_{n+1}\phi ^{\prime }\left(
t_{n+1}\right) }{t_n\phi ^{\prime }\left( t_n\right) -l_n}=t_{n+1}\phi
^{\prime }\left( t_{n+1}\right) \left[ 1-\frac{\phi ^{\prime }\left(
t_{n+1}\right) }{\phi ^{\prime }\left( t_n\right) }\right] 
\,,  
\label{120}
\end{equation}
where Eq.~(\ref{110}) was used.

Eqs. (\ref{110}) and (\ref{120}) define the 2-$d$ mapping $\left( t_n,l_n\right)
\rightarrow \left( t_{n+1},l_{n+1}\right) $, where $t_n$ is related to the capacity
of the $n$-th component $m_n$ by Eq.~(\ref{100}). This mapping can be reduced to a
1-$d$ one, because the ``first integral'' of the 2-$d$ mapping can easily be found. Namely, using
Eq.~(\ref{120}), one can express $l_n$ through $t_n$ and $t_{n-1}$, and
substitute this expression into Eq.~(\ref{110}). The result is  
\begin{equation}
\frac{t_{n+1}}{\phi ^{\prime }\left( t_n\right) }=\frac{t_n}{\phi ^{\prime
}\left( t_{n-1}\right) }
\,.  
\label{130}
\end{equation}
Repeatedly applying this relation and taking into account the fact that the
(limiting) starting point of $\left( t_n,l_n\right) $ sequence is $\left(
1,0\right) $, we obtain the following recursion relation: 
\begin{equation}
t_{n+1}=\frac{\phi ^{\prime }\left( t_n\right) }{\phi ^{\prime }\left(
1\right) }=\phi _1\left( t_n\right) 
\,.  
\label{140}
\end{equation}

\section{Sewing together}
\label{sawing}

In the last two sections we described the recursion relations in the problem under consideration in two limiting cases. Now we must sew them together.   

Let us define $G_l\left( t\right) $ through the recursion relation $%
G_{l+1}\left( t\right) =\phi _1\left[ G_l\left( t\right) \right] $ with the
initial condition $G_0\left( t\right) =t$. 
Introducing $f_l\left(
x\right) $ as $G_l\left( t\right) =f_l\left( z_1^l\ln \frac 1t\right) $, we
have $f_{l+1}\left( x\right) =\phi _1\left[ f_l\left( x/z_1\right) \right] $%
, $f_0\left( x\right) =e^{-x}$, which exactly coincides with Eq.~(\ref{38}). Then in
the limit $l\rightarrow \infty $ we have $G_l\left( t\right) =f\left(
z_1^l\ln \frac 1t\right) $, where $f\left( x\right) =\lim_{l\rightarrow
\infty }f_l\left( x\right) $ must be found from Eq.~(\ref{40}). This is the
same function as that in Eq.~(\ref{50}).

Iterating Eq.~(\ref{140}) $l\rightarrow \infty $ times yields 
\begin{equation}
t_{n+l}=f\left( z_1^l\ln \frac 1{t_n}\right) =f\left[ z_1^l\left(
1-t_n\right) \right] 
\,.  
\label{150}
\end{equation}
Here, it must be assumed $1-t_n\rightarrow 0$ together with $l\rightarrow \infty $. Distribution functions $P_n(t_n)$ and $P_{n+l}(t_{n+l})$ are
connected by the relation $P_{n+l}\left( t_{n+l}\right) dt_{n+l}=P_n\left( t_n\right)
dt_n $. Therefore, we obtain  
\begin{equation}
P_{n+l}\left( t\right) =z_1^{-l}\left| f^{-1\prime }\left( t\right) \right|
P_n\left[ 1-z_1^{-l}f^{-1}\left( t\right) \right] 
\,,  
\label{160}
\end{equation}
where $f^{-1}$ and $f^{-1\prime }$ are an inverse function and its derivative.

As $n\rightarrow \infty $ and $N\rightarrow \infty $ under the condition $%
z_1^{-n}N\rightarrow 0$, the distribution of the capacity of the $n$-th
connected component $m_n=M_n/N$ can be obtained from Eq.~(\ref{50}). But in
this limit we have from Eq.~(\ref{100}): $m_n=M_n/N=1-\phi \left( t_n\right)
=z_0\left( 1-t_n\right) $, $z_0\equiv \phi ^{\prime }\left( 1\right) $.
Then, we obtain:  
\begin{equation}
P_n\left( t\right) =z_0\left( z_1-1\right) z_1^{-n}Np\left[ z_0\left(
z_1-1\right) z_1^{-n}N\left( 1-t\right) \right] 
\,.  
\label{170}
\end{equation}
Substituting Eq.~(\ref{170}) into Eq.~(\ref{160}) and denoting $n+l$ as $n$ 
yields finally:  
\begin{equation}
P_n\left( t\right) =\nu _n\left| f^{-1\prime }\left( t\right) \right|
p\left[ \nu _nf^{-1}\left( t\right) \right] \,;\;\nu _n=z_0\left(
z_1-1\right) z_1^{-n}N
\,.  
\label{180}
\end{equation}
This formula is valid if $N\rightarrow \infty $, $n\rightarrow \infty $ 
without any restriction on the order of the limits.

\section{General results}
\label{isum}

Thus, we suggest a regular procedure for calculating the statistical properties of
intervertex distances in a random network. This procedure is valid for any large graph
with uncorrelated vertices, provided that the degree distribution $%
\Pi \left( q\right) $ has finite first and second moments. Quantities of
interest may be expressed in terms of the function $g\left( y\right) $ and its
inverse Laplace transform $p\left( x\right) $. To obtain them one has to
perform the following steps:

\begin{enumerate}

\item  \label{st1}
Calculate the $Z$-transform of the distribution function $%
\phi \left( x\right) $, Eq. (\ref{10}), and $\phi _1\left( x\right) \equiv \phi
^{\prime }\left( x\right) /\phi ^{\prime }\left( 1\right) $. 

\item  \label{st2}
Find $f\left( y\right) $, which is the solution of the
equation $f\left( z_1y\right) =\phi _1\left[ f\left( y\right) \right] $,
where $z_1=\phi _1^{\prime }\left( 1\right) $, with the conditions $f\left(
0\right) =1$, $f^{\prime }\left( 0\right) =-1$. 

\item  \label{st3}
Obtain $g\left( y\right) =\phi \left[ f\left( y\right)
\right] $. 

\item  \label{st4}
Calculate $p\left( x\right) $, which is the inverse
Laplace transform of $g\left( y\right) $, Eq. (\ref{50}).

\end{enumerate}

The most nontrivial is step \ref{st2}---no general methods for the
analytic solution of such functional equations are known. However,
asymptotic behaviour of $f\left( y\right) $ may be easily extracted. At $%
y=0$, we have $f\left( y\right) =1-y+o\left( y\right) $. Therefore, $g\left(
y\right) =1-z_0y+o\left( y\right) $, $z_0=\phi ^{\prime }\left( 1\right) $.
As $y\rightarrow +\infty $, we have $f\left( y\right) \rightarrow t_c$,
where $0\le t_c<1$ is the root of the equation $t_c=\phi _1\left( t_c\right) 
$. At large positive $y$, one can write $f\left( y\right) =t_c+h\left(
y\right) $, where $h\left( y\right) \rightarrow 0$ when $y\rightarrow
+\infty $. Then Eq. (\ref{40}) can be linearized with respect to $h$, which
gives: 
\begin{equation}
h\left( z_1y\right) =z_ch\left( y\right) 
\,,  
\label{1815}
\end{equation}
where $z_c=\phi _1^{\prime }\left( t_c\right) <1$. Looking for the solution in
the form $h\left( y\right) =Ay^{-\alpha }$, one can easily obtain the exponent  $\alpha
=-\ln z_c/\ln z_1>0$. Then we obtain asymptotic behaviour of $g\left(
y\right) $ at large positive $y$: 
\begin{equation}
g\left( y\right) =\phi \left( t_c\right) +\tilde{g}\left( y\right)
=1-m_\infty +m_\infty \tilde{g}\left( y\right) \,,\;\tilde{g}\left( y\right)
\approx By^{-\alpha }\,,\;\alpha =\frac{\ln \left( 1/z_c\right) }{\ln z_1}%
\,.\,  
\label{182}
\end{equation}

For $p\left( x\right) $, the inverse Laplace transform of $g\left( y\right) $, we
have: 
\begin{equation}
\int_0^\infty dx\,p\left( x\right) =g\left( 0\right) =1\,,\;\int_0^\infty
dx\,xp\left( x\right) =g^{\prime }\left( 0\right) =z_0=\phi ^{\prime }\left(
1\right) 
\,.  
\label{184}
\end{equation}
From $g\left( +\infty \right) =1-m_\infty $ it follows that $p\left(
x\right) $ has a $\delta $-functional part, $p\left( x\right) =\left(
1-m_\infty \right) \delta \left( x\right) +m_\infty \tilde{p}\left( x\right) 
$, $\tilde{p}\left( x\right) $ and $\tilde{g}\left( y\right) $ are related
through the Laplace transform. From the asymptotic expression for $\tilde{g%
}\left( y\right) $ at large $y$ it follows the one for $\tilde{p}\left(
x\right) $ at small $x$: 
\begin{equation}
\tilde{p}\left( x\right) \cong \frac B{\Gamma \left( \alpha \right) }%
x^{\alpha -1}
\,.  
\label{186}
\end{equation}

Various physical quantities may be expressed in terms of the functions $g(x)$ and $p(x)$. For example, the distribution functions ${\cal P}_n\left(
m\right) $ of the relative size of the $n$-th connected component, $m_n=M_n/N$
can be expressed from Eq. (\ref{180}) and the relation \thinspace $%
m_n=1-\phi \left( t_n\right) $. We have: 
\begin{equation}
{\cal P}_n\left( m\right) =\nu _n\left| g^{-1\prime }\left( 1-m\right)
\right| p\left[ \nu _ng^{-1}\left( 1-m\right) \right] 
\,,  
\label{188}
\end{equation}
where $g^{-1}$ is the function, inverse to $g(x)$. One can write ${\cal P}%
_n\left( m\right) =\left( 1-m_\infty \right) \delta \left( m\right)
+m_\infty \tilde{\cal{P}}_n\left( m\right) $, where the first term
corresponds to finite connected components of the graph, and the second one---to the giant connected component. The order of a connected component, $n$, and the
size of the graph, $N$, enter in the distribution functions in the 
combination $\nu_n$, which may be written as  
\begin{equation}
\nu _n=z_1^{n_0-n}\,,\;n_0=\frac{\ln \left[ z_0\left( z_1-1\right) N\right] 
}{\ln z_1}
\,.  
\label{189}
\end{equation}
Therefore, ${\cal P}_n\left( m,N\right) ={\cal P}\left( n-n_0,m\right) $. If 
$n<n_0$, $n_0-n\gg 1$, one can write 
\begin{equation}
{\cal P}\left( n-n_0,m\right) \cong z_1^{n_0-n}p\left( z_1^{n_0-n}m\right)
\,.  
\label{1893}
\end{equation}
This limit corresponds to the sizes of connected components being infinitely
small compared with the graph size. In this case, for $m\ll
z_1^{n-n_0}$ the small-size asymptotics of the distribution function is  
\begin{equation}
{\cal P}\left( n-n_0,m\right) \cong \frac B{\Gamma \left( \alpha \right) }%
z_c^{n-n_0}m^{\alpha -1}
\,.  
\label{1894}
\end{equation}
In the opposite limit, $n>n_0$, $n-n_0\gg 1$, the contribution of the giant connected component
to the distribution function is concentrated near $m=m_\infty $. Here
one can write, inverting the asymptotic expression (\ref{182}) for $g$: 
\begin{equation}
{\cal P}\left( n-n_0,m\right) \cong \frac{z_1^{n_0-n}}{\alpha B}\left( 
\frac{m_\infty -m}B\right) ^{-1/\alpha -1}p\left[ z_1^{n_0-n}\left( \frac{%
m_\infty -m}B\right) ^{-1/\alpha }\right] 
\, . 
\label{1895}
\end{equation}
In this case, for $z_c^{n-n_0}\ll m_\infty -m\ll 1$, we have: 
\begin{equation}
{\cal P}\left( n-n_0,m\right) \cong \frac{z_c^{n-n_0}}{\Gamma \left(
\alpha +1\right) }\left( \frac B{m_\infty -m}\right) ^2
\,.  
\label{1897}
\end{equation}

Let $Q_n$ be the probability that two randomly chosen vertices are separated
by a distance less than or equal to $n$. In fact, this is a function of $\nu _n$%
, or, equivalently, of $n-n_0$, see Eq. (\ref{189}). That is, we have $Q_n=%
{\cal Q}\left( n-n_0\right) $. This hull function, ${\cal Q}\left( l\right) $%
, can be expressed as  
\begin{equation}
{\cal Q}\left( l\right) =\int_0^{m_\infty }dm\,m{\cal P}\left( l,m\right)
=\int_0^\infty dx\,p\left( x\right) \left[ 1-g\left( z_1^lx\right) \right]
\,.  
\label{190}
\end{equation}
Here we used Eq. (\ref{188}) and introduced $x=z_1^{-l}g^{-1}\left(
1-m\right) $ as the integration variable. At $l<0$, $\left| l\right| \gg 1$, 
we have $g\left( z_1^lx\right) \cong
1-z_0z_1^lx$ in the actual region of integration. Then, taking into account Eq. (\ref{184}), we obtain  
\begin{equation}
{\cal Q}\left( l\right) \cong z_0^2z_1^l
\,.  
\label{191}
\end{equation}
As $l=n-n_0\rightarrow +\infty $, the multiple in the square brackets in the
integral in Eq. (\ref{190}) becomes equal to $m_\infty $ everywhere except
at $x=0$, where this multiple is zero. Then we have: 
\begin{equation}
\lim_{l\rightarrow +\infty }{\cal Q}\left( l\right) =m_\infty
\int_{+0}^\infty dx\,p\left( x\right) =m_\infty ^2
\,.  
\label{192}
\end{equation}
Here the $\delta $-functional part of $p\left( x\right) $ is excluded from the
integral. This result is obvious---the distance between two vertices is less
than infinity if both the vertices belong to the giant connected component. One can show
(see Appendix \ref{q}) that for large positive $l$,  
\begin{equation}
{\cal Q}\left( l\right) \cong m_\infty ^2\left[ 1-\frac{B^2\ln \left(
1/z_c\right) }{\Gamma \left( \alpha +1\right) }\left( l-l_0\right)
z_c^l\right] 
\,\,,  
\label{193}
\end{equation}
where 
\begin{eqnarray}
l_0 &=&\frac{\Gamma \left( \alpha +1\right) }{B\ln \left( 1/z_c\right) }%
\int_0^\infty dx\,x^{-\alpha }\left[ 2\ln x-\psi \left( \alpha \right)
\right] \left[ x\tilde{p}^{\prime }\left( x\right) -\left( \alpha -1\right) 
\tilde{p}\left( x\right) \right] =  
\nonumber 
\\[5pt]
&=&\frac \alpha {B\ln \left( 1/z_c\right) }\int_0^\infty dy\,y^{\alpha
-1}\left[ 2\ln y-\psi \left( \alpha \right) \right] \left[ y\tilde{g}%
^{\prime }\left( y\right) +\alpha \tilde{g}\left( y\right) \right] 
\,,
\label{194}
\end{eqnarray}
where $\psi \left( \alpha \right) =\Gamma ^{\prime }\left( \alpha \right)
/\Gamma \left( \alpha \right) $.

It is convenient to characterize the distribution of distances in the
graph using the size-independent (in the thermodynamic limit) probability
density of $l=n-n_0$, $R\left( l\right) $: 
\begin{equation}
R\left( l\right) =\frac 1{m_\infty ^2}\frac{d{\cal Q}\left( l\right) }{dl}
\, .
\label{195}
\end{equation} 

The average value and the dispersion of $l$ are equal to    
\begin{equation}
\bar{l}=-\frac 1{\ln z_1}\left[ 2\int_0^\infty dx\,\tilde{p}\left( x\right)
\ln x+\gamma _e\right] =-\frac 1{\ln z_1}\left[ 2\int_0^\infty dy\,\tilde{g}%
^{\prime }\left( y\right) \ln y-\gamma _e\right] 
\,,  
\label{196}
\end{equation}
\begin{eqnarray}
\overline{\left( l-\bar{l}\right) ^2} &=&\frac 1{\ln ^2z_1}\left\{
2\int_0^\infty dx\,\tilde{p}\left( x\right) \ln ^2x-2\left[ \int_0^\infty
dx\,\tilde{p}\left( x\right) \ln x\right] ^2+\frac{\pi ^2}6\right\} = 
\nonumber 
\\[5pt]
&=&-\frac 1{\ln ^2z_1}\left\{ 2\int_0^\infty dy\,\tilde{g}^{\prime }\left(
y\right) \ln ^2y+2\left[ \int_0^\infty dy\,\tilde{g}^{\prime }\left(
y\right) \ln y\right] ^2+\frac{\pi ^2}6\right\}  
\label{197}
\end{eqnarray}
(see Appendix \ref{q}). $\gamma _e=0.577216\dots $ is the Euler-Masceroni constant. 

Note that the asymptotic formulae (\ref{182}), (\ref{186}), (\ref{1894}%
), (\ref{1895}), (\ref{1897}) and (\ref{193}) are valid only if $\phi
^{\prime }\left( t_c\right) \ne 0$ and $\phi _1^{\prime }\left( t_c\right)
\sim \phi ^{\prime \prime }\left( t_c\right) \ne 0$. This conditions are
violated if $\phi ^{\prime }\left( 0\right) =0$, which is the case when $t_c=0$ and $%
m_\infty =1$. If $\phi ^{\prime }\left( 0\right) =0$, but $\phi _1^{\prime
}\left( 0\right) \ne 0$ (vertices of degree one are absent, but vertices
of degree two are present), the asymptotics of
 $f\left( y\right) $ at large $y$
is again $f\left( y\right) \cong Ay^{-\alpha }$, $\alpha =\ln \left(
1/z_c\right) /\ln z_1$, $z_c=\phi _1^{\prime }\left( 0\right) =2\Pi \left(
0\right) /\bar{q}$. In this particular case, because of $g\left( y\right) =\phi \left[ f\left(
y\right) \right] \sim \left[ f\left( y\right) \right] ^2\sim y^{-2\alpha }$,
in all formulae $\alpha$ must be replaced with $2\alpha $, and,
respectively, $z_c$ with $z_c^2$. 

The situation is different if $\phi _1^{\prime }\left( 0\right) =\phi
_1^{\prime \prime }\left( 0\right) =0$. Assume that the minimal vertex degree in
the network is $k\ge 3$. Then at $x\rightarrow \infty $, $\phi \left(
x\right) \cong \Pi \left( k\right) x^k$ and $\phi _1\left( x\right)
\cong k\Pi \left( k\right) x^{k-1}/\bar{q}$. So, instead of Eq. (\ref
{1815}), we have at large $y$: 
\begin{equation}
f\left( y\right) =\zeta _k\left[ f\left( y/z_1\right) \right]
^{k-1}\,,\;\zeta _k=\frac{k\Pi \left( k\right) }{\bar{q}}\le 1
\,,
\label{1971}
\end{equation}
where $\zeta _k=1$ only if $\Pi \left( q\right) =\delta _K\left( q-k\right) $
(in this case the equation for $f$ can be solved exactly). Its
solution is  
\begin{equation}
f\left( y\right) =\zeta _k^{-1/\left( k-2\right) }\exp \left( -Ay^b
\right) 
\,,\;
b =\frac{\ln \left( k-1\right) }{\ln z_1}<1
\,,  
\label{1792}
\end{equation}
with some constant positive coefficient $A$. Let us prove that $b \le 1$, i.e. $z_1\ge
k-1 $. Indeed,  
\begin{equation}
\bar{q}z_1-\bar{q}\left( k-1\right) =\sum_{q=k}^\infty q\left( q-k\right)
\Pi \left( q\right) \ge 0\,.  \label{1973}
\end{equation}
The equality is possible only in the case $\Pi \left( q\right) =\delta
_K\left( q-k\right) $. The asymptotics of $g\left( y\right) $ at large $y$
is  
\begin{equation}
g\left( y\right) \cong \Pi \left( k\right) \left[ f\left( y\right) \right]
^k\cong \left( \frac{\bar{q}}k\right) ^{k/\left( k-2\right) }\left[ \Pi
\left( k\right) \right] ^{-2/\left( k-2\right) }\exp \left( -By^b
\right) 
\,,  
\label{1974}
\end{equation}
where $B=kA$. The asymptotics of the function $p\left( x\right) $ at $%
x\rightarrow 0$ can be obtained by making the saddle point evaluation of the
inverse Laplace transform integral in Eq. (\ref{50}). We have: 
\begin{eqnarray}
&&p\left( x\right) \cong Dx^{-\left( 2-b \right) /\left[ 2\left(
1-b \right) \right] }\exp \left( -Cx^{-b /\left( 1-b \right)
}\right) \,,  \nonumber \\
&&C=\left( b ^{b /\left( 1-b \right) }-b ^{1/\left( 1-b
\right) }\right) B^{1/\left( 1-b \right) }\,,\;D=\frac{\left( b
B\right) ^{1/\left[ 2\left( 1-b \right) \right] }}{\sqrt{2\pi \left(
1-b \right) }}
\,,  
\label{1975}
\end{eqnarray}
The asymptotic expressions (\ref{1894})--(\ref{1897}) must be replaced with a
new ones in this case. For brevity, we present here explicitly only the
asymptotic expression for the cumulative distance distribution ${\cal Q}\left(
l\right) $ when the distance deviation $l$ is positive and large, i.e. the
one which replaces Eq. (\ref{193}). 
We have: 
\begin{eqnarray}
& & {\cal Q}\left( l\right) =1-\int_0^\infty dx\,p\left( x\right) g\left( x/\nu
\right) \cong 
\nonumber
\\[5pt]
& & 
1-F\int_0^\infty dx\,x^{-\left( 2-b \right) /\left[
2\left( 1-b \right) \right] }\exp \left( -Cx^{-b /\left( 1-b
\right) }-B\nu ^{-b }x^b \right) 
\,,  
\label{1976}
\end{eqnarray}
assuming that the actual region of integration is $\nu \ll x\ll 1$. Here $%
F=\left( \bar{q}/k\right) ^{k/\left( k-2\right) }\left[ \Pi \left( k\right)
\right] ^{-2/\left( k-2\right) }D$ and $\nu =z_1^l$. Then, the saddle point
calculation of this integral gives: 
\begin{equation}
{\cal Q}\left( l\right) \cong 1-Gz_1^{l/2}\exp \left[ -Hz_1^{l/\left(
2-b \right) }\right] 
\,,  
\label{1977}
\end{equation}
where $G$ and $H$ are some positive numbers which can be expressed in terms of  $b $, 
$B$, $k$ and $\Pi \left( k\right) $.

\section{Networks with an exponentially decaying degree distribution}
\label{exp}

Here we present a two-parameter family of degree distributions, for which one
can obtain exact analytical expressions for ${\cal P}_n\left( t\right) $,
and, consequently, for the intervertex distance distribution. These are the degree distributions, for
which $\phi _1\left( x\right) $ is a fractional linear function. These
functions form a group with respect to the operation of functional
composition. Indeed, the composition of any fractional linear functions is a fractional linear
function, and the inverse of any fractional linear function is also a
fractional linear function. Then, one can look for the solution of Eq.~(\ref{40}), 
$f\left( y\right) $, in the form of a fractional linear function too. It is
more convenient to define a one-parameter family of linear fractionals $%
f\left( x\right) $, and then to write $\phi _1$ as $\phi _1\left( x\right)
=f\left[ z_1f^{-1}\left( x\right) \right] $.

Any linear fractional $f\left( x\right) $, under the conditions $f\left(
0\right) =1$ and $f^{\prime }\left( 1\right) =-1$ may be written as  
\begin{equation}
f\left( y\right) =t_c+\frac{\left( 1-t_c\right) ^2}{1-t_c+y}
\,,  
\label{200}
\end{equation}
It depends on one parameter $t_c=f\left( \infty \right) $, whose meaningful
values belong to the interval $\left( 0,1\right) $. Then $\phi _1(x)$ may be
expressed as  
\begin{equation}
\phi _1\left( x\right) =\frac{\left( z_1-1\right) t_c+\left( 1-z_1t_c\right)
x}{z_1-t_c-\left( z_1-1\right) x}
\,.  
\label{210}
\end{equation}
The degree distribution $\phi \left( x\right) \sim \int dx\,\phi
_1\left( x\right) $, $\phi \left( 1\right) =1$, can be restored up to an 
additive constant. Since $\phi \left( 0\right) $ is the fraction of
zero-degree vertices, whose effect on the properties of the network is
trivial, it is natural to set the integration constant so that such
vertices will be excluded, $\phi \left( 0\right) =0$. The result is 
\begin{equation}
\phi \left( x\right) =\beta x+\left( 1-\beta \right) \frac{\ln \left(
1-\zeta x\right) +\zeta x}{\ln \left( 1-\zeta \right) +\zeta }
\,,
\label{220}
\end{equation}
where the parameters $\beta $ and $\zeta $ are connected with $z_1$ and $t_c$ as
follows: 
\begin{eqnarray}
z_1 &=&\frac{\left( 1-\beta \right) \zeta ^2}{1-\zeta }\frac 1{\zeta
^2-\beta \left[ \left( 1-\zeta \right) \ln \left( 1-\zeta \right) +\zeta
\right] }
\,,  
\nonumber 
\\[5pt]
t_c &=&\beta \frac{1-\zeta }\zeta \frac{\ln \left( 1-\zeta \right) +\zeta }{%
\beta \left[ \left( 1-\zeta \right) \ln \left( 1-\zeta \right) +\zeta
\right] -\zeta ^2}
\,.  
\label{227}
\end{eqnarray}

The degree distribution in the original representation is  
\begin{equation}
\Pi \left( q\right) =\beta \delta \left( q-1\right) -\left[ 1-\delta \left(
q-1\right) \right] \frac{1-\beta }{\ln \left( 1-\zeta \right) }\frac{\zeta ^q%
}q
\,.  
\label{225}
\end{equation}
In this case, the average vertex degree is  
\begin{equation}
\bar{q} \equiv z_0=\phi ^{\prime }\left( 1\right) =\beta -\frac{1-\beta }{\left(
1-\zeta \right) \ln \left( 1-\zeta \right) }
\,\,   
\label{230}
\end{equation}
and the relative size of the giant connected component is  
\begin{equation}
m_\infty =1-\phi \left( t_c\right) =\frac{\left( 1-t_c\right) \left[ \beta
\ln \left( 1-\zeta \right) +\zeta \right] -(1-\beta )\ln z_1}{\ln \left(
1-\zeta \right) +\zeta }
\,.  
\label{240}
\end{equation}


\begin{figure}
\epsfxsize=60mm
\centerline{\epsffile{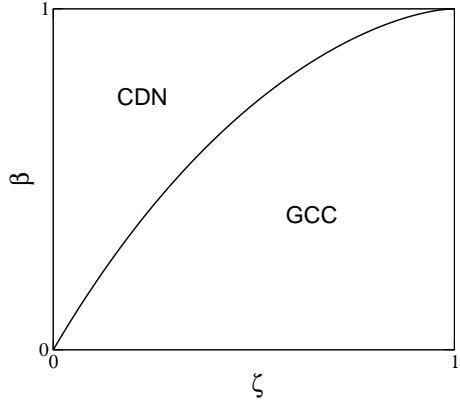}}
\caption{
Phase diagram of the model with an exponentially decaying degree
distribution. Here the GCC indicates the presence of a giant connected
component in the network, and the CDN---a completely disconnected network. 
The capacity (relative size) of the giant connected component is $%
m_\infty=1$ along the line $\beta=0$.
}
\label{phdia}
\end{figure}


The giant connected component exists if $z_1>1$, which in our case corresponds
to $\beta <\beta _c\left( \zeta \right) $, where  
\begin{equation}
\beta_c(\zeta)=
\frac{\zeta ^3}{\zeta \left( 2\zeta -1\right) -\left( 1-\zeta
\right) ^2\ln \left( 1-\zeta \right) }
\,.  
\label{245}
\end{equation}
The phase diagram of the model is shown in fig. \ref{phdia}. It should be noted,
that the giant connected component disappears if the number of
one-degree vertices (dead ends) exceeds some critical value. If these
vertices are absent, (almost) the entire network is a single connected component. 

\begin{figure}
\epsfxsize=100mm
\centerline{\epsffile{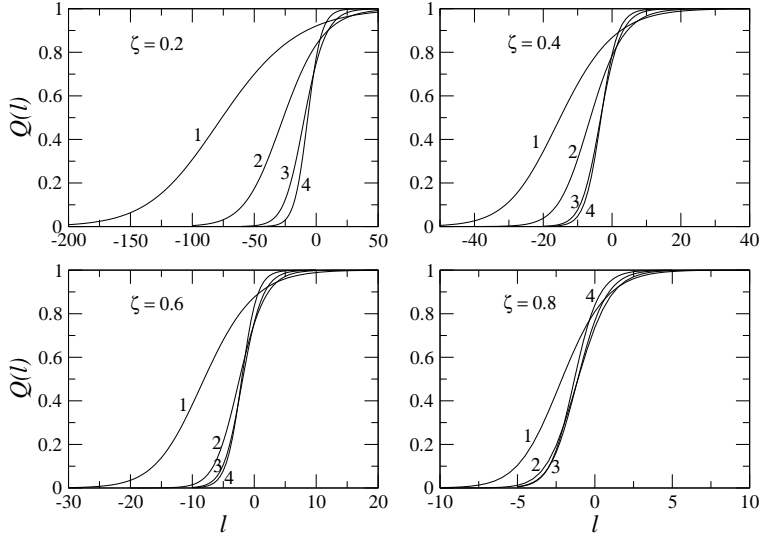}}
\caption{
Cumulative distance distribution function ${\cal Q}(l)$ in the model with
an exponentially decaying degree distribution for various values of $\zeta$ and $\beta$. 
When $\zeta=0.2$, curves 1, 2, 3, and 4 correspond to 
$\beta=0.3, 0.25, 0.15$, and $0.05$, respectively. 
If $\zeta=0.4$, curves 1, 2, 3, and 4 correspond to 
$\beta=0.5,\ 0.4,\ 0.2,\ 0.1$, respectively. 
If $\zeta=0.6$, curves 1, 2, 3, and 4 correspond to 
$\beta=0.7,\ 0.5,\ 0.3,\ 0.1$, respectively. 
If $\zeta=0.8$, curves 1, 2, 3, and 4 correspond to 
$\beta=0.8,\ 0.6,\ 0.4,\ 0.2$, respectively. 
}
\label{q-exp}
\end{figure}

The composition function $g\left( x\right) \equiv \phi \left[ f\left( x\right)
\right] $, which is the Laplace transform of the distribution function $%
p\left( s\right) $ of $s_n=z_0\left( z_1-1\right) z_1^{-n}M_n$, $M_n$ being
the size of the $n$-th connected component for a large but finite $n$, is 
\begin{equation}
g\left( y\right) =1-m_\infty +\frac{\beta \left( 1-t_c\right) ^2}{1-t_c+y}+%
\frac{1-\beta }{\ln \left( 1-\zeta \right) +\zeta }\left\{ \ln \left[ \frac{%
\left( 1-t_c\right) /z_1+y}{1-t_c+y}\right] +\frac{\zeta \left( 1-t_c\right)
^2}{1-t_c+y}\right\} 
\,.  
\label{250}
\end{equation}
The calculation of the inverse Laplace transform is straightforward: 
\begin{eqnarray}
& & p\left( x\right) = \left( 1-m_\infty \right) \delta \left( x\right) +\beta
\left( 1-t_c\right) ^2\exp \left[ -\left( 1-t_c\right) x\right] -  
\nonumber
\\[5pt] 
&&\frac{1-\beta }{\ln \left( 1-\zeta \right) +\zeta }\left\{ \frac 1x\left[
\exp \left[ -\left( 1-t_c\right) x/z_1\right] -\exp \left[ -\left(
1-t_c\right) x\right] \right] -\zeta \left( 1-t_c\right) ^2\exp \left[
-\left( 1-t_c\right) x\right] \right\} 
.  
\label{260}
\end{eqnarray}
Below, for the sake of simplicity, we present the
results for $\beta =0$ only, when $\Pi \left( 1\right) =0$ and the ``dead ends''
are absent. 
(See results for $b \neq 0$ in fig. \ref{q-exp}.)  
In this case $t_c=0$ and $m_\infty =1$, i.e. the giant connected
component (almost) coincides with the whole graph. 


The distribution function ${\cal P}_n\left( m\right) ={\cal P}\left(
n-n_0,m\right) $ depends upon the size of the network through $n_0$: 
\begin{equation}
n_0=1+\ln \left| \frac{\ln \left( 1-\zeta \right) +\zeta }{\zeta ^3}\right|
\,.  
\label{2601}
\end{equation}
Its dependence on $m$ can be represented in a parametric form  
by introducing
a parameter $t$, related to the size of the connected component $m$ as  
\begin{equation}
m\left( t\right) =1-\phi \left( t\right) =\frac{\ln \left[ \left( 1-\zeta
t\right) /\left( 1-\zeta \right) \right] -\zeta \left( 1-t\right) }{\left|
\ln \left( 1-\zeta \right) +\zeta \right| }  
\, . 
\label{2602}
\end{equation}
Using this parametrization, Eq. (\ref{188}) can be written as: 
\begin{eqnarray}
&&{\cal P}\left( l,t\right) =\nu \left| \frac{f^{-1\prime }\left( t\right) }{%
m^{\prime }\left( t\right) }\right| p\left[ \nu f^{-1}\left( t\right)
\right] =  
\nonumber 
\\[5pt]
&&\frac{1-\zeta t}{\zeta ^2t^2\left( 1-t\right) }\left\{ \exp \left[
-\left( 1-\zeta \right) \nu \frac{1-t}t\right] -\left( 1+\zeta \nu \frac{1-t}%
t\right) \exp \left( -\nu \frac{1-t}t\right) \right\} ,\;\nu =\left( 1-\zeta
\right) ^l
.  
\label{2603}
\end{eqnarray}
Eqs. (\ref{2602}) and (\ref{2603}) determine ${\cal P}\left( l,m\right) $ in
a parametric form. At small $m$, ${\cal P}\left( l,m\right) $ has the asymptotics: 
\begin{equation}
{\cal P}\left( l,m\right) \cong \frac{\left( 1-\zeta \right) \bar{q}}{%
\zeta ^2m}\left[ \exp \left( -\left( 1-\zeta \right) \nu \frac m{\bar{q}}%
\right) -\left( 1+\zeta \nu \frac m{\bar{q}}\right) \exp \left( -\nu \frac m{%
\bar{q}}\right) \right] 
\,,  
\label{2604}
\end{equation}
where $\bar{q}=\phi ^{\prime }\left( 1\right) =\zeta ^2/\left[ \left(
1-\zeta \right) \left| \ln \left( 1-\zeta \right) +\zeta \right| \right] $
is the average vertex degree. Equation (\ref{2605}) is valid if $m\ll 1$. On the
other hand, as $1-m\ll 1,\nu $, we have: 
\begin{equation}
{\cal P}\left( l,m\right) \cong \frac{1-\zeta }{2\zeta ^2\left( 1-m\right) 
}\exp \left[ -\left( 1-\zeta \right) \nu \left( \sqrt{\frac{1-\zeta }{%
2\left( 1-m\right) }}-1\right) \right] \,.  \label{2605}
\end{equation}

The cumulative distance distribution $Q_n$ is a function of $%
l=n-n_0\left( N\right) $: 
\begin{equation}
{\cal Q}\left( l\right) =\int_0^1dt\,\left| m^{\prime }\left( t\right)
\right| m\left( t\right) {\cal P}\left( l,t\right) 
\,.  
\label{2606}
\end{equation}
We failed to find this integral analytically, but asymptotic 
expressions can be presented. As $l>0$ and $l\gg 1$,  
\begin{equation}
{\cal Q}\left( l\right) \cong 1-A\left( l-l_0\right) \left( 1-\zeta
\right) ^{2l}
\,,  
\label{2607}
\end{equation}
\begin{equation}
A=\left[ \frac{\zeta ^2}{2\left[ \zeta +\ln \left( 1-\zeta \right) \right] }%
\right] ^2\,,\;l_0=\frac{\gamma _e}{\left| \ln \left( 1-\zeta \right)
\right| }-2\left( 1-\zeta \right) \frac{\zeta +\ln \left( 1-\zeta \right) }{%
\zeta ^2\ln \left( 1-\zeta \right) }
\,.  
\label{2608}
\end{equation}
Here we have $\left( 1-\zeta \right) ^{2l}$ instead of $\left( 1-\zeta
\right) ^l$ in the asymptotics, because one-degree vertices are absent in
the network, but vertices of degree two are present. On the other hand, as $%
l<0 $ and $\left| l\right| \gg 1$, we have  
\begin{equation}
{\cal Q}\left( l\right) \cong \left( \frac{\zeta ^2}{\ln \left( 1-\zeta
\right) +\zeta }\right) ^2\left( 1-\zeta \right) ^{-l-2}
\,.  
\label{2609}
\end{equation} 

The position of the center of the distance distribution and its mean square
deviation are given by Eqs. (\ref{196}) and (\ref{197}) respectively. Calculating
the integrals yields  
\begin{equation}
\bar{l}=\frac{\gamma _e}{\left| \ln \left( 1-\zeta \right) \right| }-\frac{%
\ln \left( 1-\zeta \right) }{\ln \left( 1-\zeta \right) +\zeta }
\,,
\label{26091}
\end{equation}
\begin{equation}
\overline{\left( l-\bar{l}\right) ^2}=\frac{\pi ^2}{2\ln ^2\left( 1-\zeta
\right) }+\frac 23\frac{\ln \left( 1-\zeta \right) }{\ln \left( 1-\zeta
\right) +\zeta }-\frac 12\left[ \frac{\ln \left( 1-\zeta \right) }{\ln
\left( 1-\zeta \right) +\zeta }\right] ^2
\,.  
\label{26092}
\end{equation}

\section{Power-law degree distribution with an exponential cut-off}
\label{power}

The general scheme, introduced in this paper, is applicable only if the
degree distribution has finite first and second moments in the thermodynamic limit. 
If, for example, the degree distribution $\Pi \left( q\right) $ is
asymptotically a ``scale-free'' one, $\Pi \left( q\right) \sim q^{-\gamma }$, 
at large $q$, and exponent $\gamma \le 3$, then our considerations fail.
In this section we shall consider the networks with power-law degree distributions, $2<\gamma <3$, and with an exponential cut-off at large degrees. 

The crucial point of our formalism is to find the general solution of the
recursive relation $t_{n+1}=\phi _1\left( t_n\right) $, or, more precisely,
to find how this solution will behave at the large number of iterations $n$. One can
see that this recursion relation is easily solvable in the following case: 
\begin{equation}
\phi _1\left( x\right) =1-\left( 1-t_c\right) \left( \frac{1-x}{1-t_c}%
\right) ^{\gamma -2}\,,  \label{2610}
\end{equation}
where $2<\gamma <3$ and $0\le t_c<1$. Indeed, we have: 
\begin{equation}
t_n=1-\left( 1-t_c\right) \left( \frac{1-t_0}{1-t_c}\right) ^{\left( \gamma
-2\right) ^n}
\,,  
\label{2620}
\end{equation}
that is the analytic form of the solution with any initial condition.
However, we have to have finite value of $z_1=\phi _1^{\prime }\left(
1\right) $ to apply the general scheme. Then, let us define  
\begin{equation}
\phi _1\left( x\right) =\frac{1-\left( 1-\zeta x\right) ^{\gamma -2}}{%
1-\left( 1-\zeta \right) ^{\gamma -2}}
\,,  
\label{2630}
\end{equation}
where we set the parameter $t_c$ regulating the size of the giant connected
component to be equal to zero, which means that the giant connected component 
contains almost all the vertices. The problem can be solved for an arbitrary $%
t_c$, but then the results would look essentially more cumbersome. The
parameter $\zeta <1$ corresponds to the cut-off. Indeed, the $Z$-transformed
degree distribution $\phi \left( x\right) $ may be easily obtained from Eq. (%
\ref{2630}), by integrating its right-hand side. We have: 
\begin{equation}
\phi \left( x\right) =\frac{\left( 1-\zeta x\right) ^{\gamma -1}-1+\zeta
\left( \gamma -1\right) x}{\left( 1-\zeta \right) ^{\gamma -1}-1+\zeta
\left( \gamma -1\right) }
\,.  
\label{2640}
\end{equation}
Then the degree distribution is  
\begin{equation}
\Pi \left( q\right) =\left[ \left( 1-\zeta \right) ^{\gamma -1}-1+\zeta
\left( \gamma -1\right) \right] ^{-1}\Gamma \left( \gamma \right) \frac{\sin
\pi \gamma }\pi \frac{\Gamma \left( q-\gamma +1\right) }{q!}\zeta ^q
\,,
\label{2650}
\end{equation}
from which one can easily see that $\Pi \left( q\right) \sim q^{-\gamma
}\zeta ^q$ at $q\rightarrow \infty $. In the following we shall assume $1-\zeta
\ll 1$.

We have to find the solution $f\left( y\right) $ of the functional equation $%
f\left( y\right) =\phi _1\left[ f\left( y/z_1\right) \right] $, or: 
\begin{equation}
1-\left[ 1-\left( 1-\zeta \right) ^{\gamma -2}\right] f\left( y\right)
=\left[ 1-\zeta f\left( y/z_1\right) \right] ^{\gamma -2}
\,,  
\label{2660}
\end{equation}
where  
\begin{equation}
z_1=\phi _1^{\prime }\left( 1\right) \cong \left( \gamma -2\right) \left(
1-\zeta \right) ^{\gamma -3}\gg 1
\,.  
\label{2670}
\end{equation}
The function $f\left( y\right) $ must satisfy the initial conditions $%
f\left( 0\right) =1$ and $f^{\prime }\left( 0\right) =-1$. Therefore, for
small enough $\left| y\right| $ we have: $f\left( y\right) \approx 1-y$.
This approximate equality holds when $\left| y\right| \ll \left| f^{\prime
}\left( 0\right) /f^{\prime \prime }\left( 0\right) \right| =1/\left|
f^{\prime \prime }\left( 0\right) \right| \sim \left( 1-\zeta \right)
^{\gamma -2}$ (see Appendix \ref{saw}). On the other hand, at large enough $%
\left| y\right| $ one can set in Eq. (\ref{2660}) $\zeta =1$ (but not $%
z_1 = \infty $!). The resulting equation  
\begin{equation}
1-f\left( y\right) =\left[ 1-f\left( y/z_1\right) \right] ^{\gamma -2}
\,,
\label{2680}
\end{equation}
can be easily solved: 
\begin{equation}
f\left( y\right) =1-\exp \left( -Ax^{-\vartheta }\right) \,,\;\frac 1\vartheta =%
\frac{\ln z_1}{\ln \left[ 1/\left( \gamma -2\right) \right] }=\frac{\left(
3-\gamma \right) \ln \left( 1-\zeta \right) }{\ln \left( \gamma -2\right) }%
+1\gg 1
\,.  
\label{2690}
\end{equation}
The constant $A$ must be determined by sewing together the expressions for
small and large $y$ (see Appendix \ref{saw}), which gives: $A=1/e\vartheta $.
Thus, we have: 
\begin{equation}
f\left( y\right) =1-\exp \left( -\frac{y^{-\vartheta }}{e\vartheta }\right) 
\,.
\label{2700}
\end{equation}
This formula is valid if $\left| y\right| \gg \left( 1-\zeta \right)
^{\gamma -2}$ (Appendix \ref{saw}), i.e. for a large enough cut-off parameter $\left| \ln
\zeta \right| $, this formula is valid almost everywhere except small vicinity of $y=0$%
. Then, $g\left( y\right) =\phi \left[ f\left( y\right) \right] $ is: 
\begin{equation}
g\left( y\right) =\frac 1{\gamma -2}\left\{ \exp \left[ -\left( \gamma
-1\right) \frac{y^{-\vartheta }}{e\vartheta }\right] -\left( \gamma -1\right) \exp
\left( -\frac{y^{-\vartheta }}{e\vartheta }\right) +\gamma -2\right\}
\,.  
\label{2710}
\end{equation}
The inverse Laplace transform of $g\left( y\right) $ is  
\begin{equation}
p\left( x\right) =\frac{\gamma -1}{\gamma -2}\frac{x^{\vartheta -1}}e\left\{
\exp \left( -\frac{x^\vartheta }{e\vartheta }\right) -\exp \left[ -\left( \gamma
-1\right) \frac{x^\vartheta }{e\vartheta }\right] \right\} 
\,   
\label{2720}
\end{equation} 
(see Appendix \ref
{saw}). 
The region of validity of this formula is $\left| x\right| \ll \left(
1-\zeta \right) ^{2-\gamma }$. 


\begin{figure}
\epsfxsize=100mm
\centerline{\epsffile{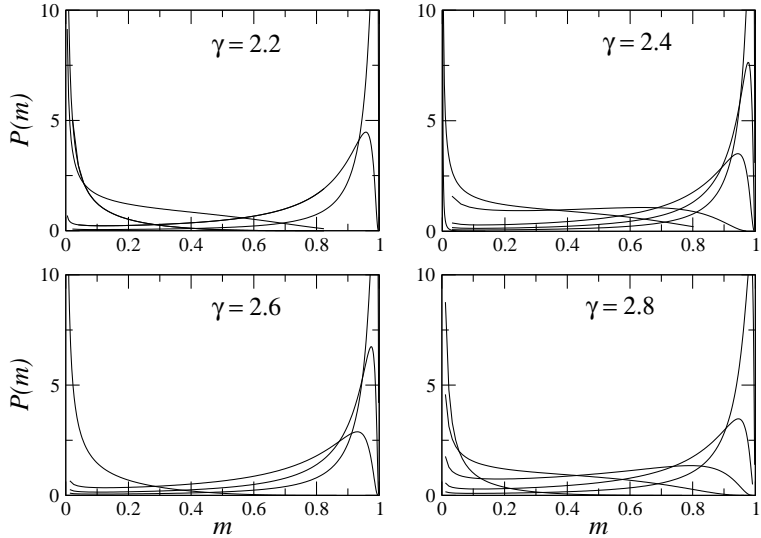}}
\caption{
Series of the connected-component size distributions 
${\cal P}(l,m)$ in the model with a power law degree distribution for various values
of the $\gamma$ exponent. 
As the distribution more and more concentrates near $m=1$, 
the parameter $l$ of the curves takes the values $0.5,\ 0,\ -1.0,\ -1.5$ when $\gamma=2.2$; 
the values $1.0,\ 0.5,\ 0,\ -1.0,\ -1.5$ when $\gamma=2.4$; 
$2.0,\ 1.0,\ 0,\ -4.0$ when $\gamma=2.6$; and  
$4.0,\ 1.0,\ -2.0,\ -5.0,\ -10.0$ when $\gamma=2.8$. 
}
\label{p-pow}
\end{figure}


Now, let us consider the distribution of the size of the $n$-th
connected component, ${\cal P}_n\left( m\right) $. Since it is impossible
to calculate the inverse of $g\left( y\right) $ analytically, we use
the parametric form of Eq. (\ref{188}), introducing a parameter $t$, which is
connected with $m$ as  
\begin{equation}
m\left( t\right) =1-\phi \left( t\right) =\frac 1{\gamma -2}\left[ \left(
\gamma -1\right) \left( 1-t\right) -\left( 1-t\right) ^{\gamma -1}\right] 
\,.
\label{2730}
\end{equation}
Then we have: 
\begin{equation}
{\cal P}_n\left( t\right) =\nu _n\left| \frac{f^{-1\prime }\left( t\right) }{%
m^{\prime }\left( t\right) }\right| p\left[ \nu _nf^{-1}\left( t\right)
\right] 
\,,  
\label{2740}
\end{equation}
where $f^{-1}\left( t\right) $ is the inverse of the function $f\left(
y\right) $, Eq. (\ref{2700}) and $\nu _n=z_0\left( z_1-1\right) z_1^{-n}N$.
Combining Eqs. (\ref{188}), (\ref{2690}), (\ref{2730}) and (\ref{2740}) we
obtain: 
\begin{equation}
{\cal P}_n\left( t\right) =\frac{\mu _n\left[ 1-\left( 1-t\right) ^{\gamma
-2}\right] ^{-1}}{\left( 1-t\right) \ln ^2\left( 1-t\right) }\left\{ \exp
\left[ \frac{\mu _n}{\ln \left( 1-t\right) }\right] -\exp \left[ \frac{%
\left( \gamma -1\right) \mu _n}{\ln \left( 1-t\right) }\right] \right\} 
\,,
\label{2750}
\end{equation}
where  
\begin{eqnarray}
\mu _n &=&\left( e\vartheta \right) ^{-2}\nu _n^\vartheta =\frac{\left( \gamma
-2\right) ^nN^\vartheta }{\left( e\vartheta \right) ^2}=\left( \gamma -2\right)
^{n-n_0}
\,,  
\label{2760} 
\\
n_0 &=&\frac{\vartheta \ln N-2\left( \ln \vartheta +1\right) }{\ln \left[ 1/\left(
\gamma -2\right) \right] }
\,,  
\label{2770}
\end{eqnarray}
$\vartheta $ is given by Eq. (\ref{2690}). 
Eqs. (\ref{2730}) and (\ref{2750})
define the distributions of the sizes of the $n$-th connected component ${\cal P}%
_n\left( m\right) $ in the parametric form, see fig. \ref{p-pow}.  
The order of a connected component, 
$n$, and the size of the system, $N$, enter here only in the combination $%
n-n_0\left( N\right) $, i.e. one can write ${\cal P}_n\left( m\right) ={\cal %
P}\left( n-n_0,m\right) $. One can write the following asymptotic
expression for ${\cal P}\left( l,m\right) $: at small $m$, $m\ll \exp
\left[ -\left( \gamma -2\right) ^l\right] $,  
\begin{equation}
{\cal P}\left( l,m\right) \cong \frac{\left( \gamma -1\right) \left(
\gamma -2\right) ^{2l}}{m\ln ^3\left( \bar{q}/m\right) }
\,,  
\label{2773}
\end{equation}
where $\bar{q}=\left( \gamma -1\right) /\left( \gamma -2\right) $ is the
mean degree. When $m$ close to $1$, or, more precisely, when $1-m\ll 1,\;\left(
\gamma -2\right) ^{2l}$, we have: 
\begin{equation}
{\cal P}\left( l,m\right) \cong \left( \gamma -2\right) ^{l-1}\left[ \frac{%
\gamma -1}{2\left( 1-m\right) }\right] ^{3/2}\exp \left\{ -\left( \gamma
-2\right) ^l\sqrt{\frac{\gamma -1}{2\left( 1-m\right) }}\right\} 
\,.
\label{2777}
\end{equation}
Obviously, the distribution is concentrated near $m=0$ when $l<0$, $\left|
l\right| \gg 1$, and near $m=1$ when $l>0$, $l\gg 1$. 

\begin{figure}
\epsfxsize=70mm
\centerline{\epsffile{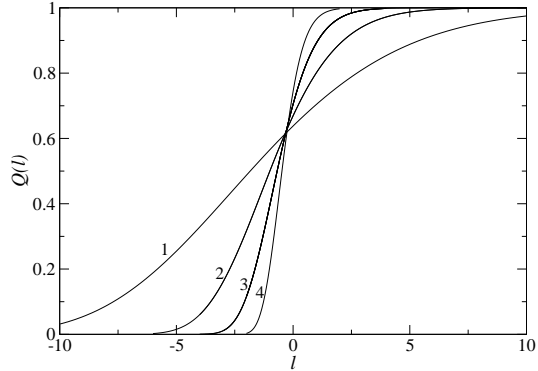}}
\caption{
Cumulative distance distribution function ${\cal Q}(l)$ in the model with a
power-law degree distribution for various values of the $\gamma$ exponent. 
Curves 1, 2, 3, and 4 correspond to $\gamma=$ 2.2, 2.4, 2.6, and 2.8, respectively. 
}
\label{q-pow}
\end{figure}

The intervertex distance distribution actually depends on $n-n_0\equiv
l $ only. For the cumulative distance distribution $Q_n={\rm Prob}\left( {\rm Distance}\le n\right) =%
{\cal Q}\left( n-n_0\right) $ we have: 
\begin{equation}
{\cal Q}\left( l\right) =\int_0^1dm\,m{\cal P}\left( l,m\right)
=\int_0^1dt\,m^{\prime }\left( t\right) {\cal P}\left( l,t\right) 
\,.
\label{2780}
\end{equation}
This integral can easily be evaluated, and we obtain  
\begin{eqnarray}
& & {\cal Q}\left( l\right) = 2\frac{\left( \gamma -1\right) \mu ^{1/2}}{\left(
\gamma -2\right) ^2}
\times 
\nonumber 
\\[5pt]
& & \left\{ \left( \gamma -1\right) K_1\left( 2\mu
^{1/2}\right) +K_1\left[ 2\left( \gamma -1\right) \mu ^{1/2}\right] -2\left(
\gamma -1\right) ^{1/2}K_1\left[ 2\left( \gamma -1\right) ^{1/2}\mu
^{1/2}\right] \right\} 
\,,  
\label{2790}
\end{eqnarray}
where $\mu =\left( \gamma -2\right) ^l$ and $K_1$ is the McDonald 
function, see fig. \ref{q-pow}. For large negative $l$, using the large argument asymptotics of $%
K_1\left( z\right) $ we obtain the following expression: 
\begin{equation}
{\cal Q}\left( l\right) \cong \left( \frac{\gamma -1}{\gamma -2}\right)
^2\pi ^{1/2}\left( \gamma -2\right) ^{l/4}\exp \left[ -2\left( \gamma
-2\right) ^{l/2}\right] 
\,.  
\label{2880}
\end{equation}

\noindent
For a large positive $l\gg 1$ we have: 
\begin{equation}
{\cal Q}\left( l\right) \cong 1-\frac 12\left( \gamma -1\right) ^2\left(
l-l_1\right) \left( \gamma -2\right) ^{2l}\ln \left( \frac 1{\gamma -2}%
\right) \,,\;l_1=\frac{2\gamma _e-3/2-2\frac{\gamma -1}{\gamma -2}\ln \left(
\gamma -1\right) }{\ln \left[ 1/\left( \gamma -2\right) \right] }
\,.
\label{2890}
\end{equation}
The hull
function ${\cal Q}\left( l\right) $ is characterized by the position of its
center: 
\begin{equation}
\bar{l}=\int_{-\infty }^{+\infty }dl\,l\frac{d{\cal Q}\left( l\right) }{dl}=-%
\frac 2{\ln \left( \gamma -2\right) }\left[ \gamma _e-\frac{\ln \left(
\gamma -1\right) }{\gamma -2}\right] 
\,   
\label{2900}
\end{equation}
and by its width  
\begin{equation}
\delta l^2=\int_{-\infty }^{+\infty }dl\,\left( l-\bar{l}\right) ^2\frac{d%
{\cal Q}\left( l\right) }{dl}=\frac 2{\ln ^2\left( \gamma -2\right) }\left[ 
\frac{\pi ^2}6-\frac{\left( \gamma -1\right) \ln ^2\left( \gamma -1\right) }{%
\left( \gamma -2\right) ^2}\right] 
\,.  
\label{2910}
\end{equation}

Note that all the results of this section were obtained assuming
two limiting transitions. 
First the size of the network tends to infinity,
while the cut-off parameter is kept finite. This allows us to apply the general
formalism based on Eq.~(\ref{40}). And only afterwards the cut-off parameter $\left|
\ln \zeta \right| $ tends to infinity. This allows us to obtain the solution
of Eq. (\ref{40}) in the leading order. The limiting transitions in this section are 
performed precisely in this order. 
Situations where these two limiting transitions must be performed simultaneously will be discussed
in the next, conclusive section.

\section{Conclusions} 
\label{concl}

The most crucial restriction in our formalism is that vertices of the network 
are uncorrelated, 
so that the network is completely defined by a given degree distribution
$\Pi \left( q\right) $. 
This allowed us to trace the evolution of the $n$-th connected component
of a vertex as the $n$ is growing. This is possible, however, only if $\Pi \left(
q\right) $ has finite first and second moments, $\bar{q}$ and $\overline{q^2}
$. These networks contain (almost) no closed loops of finite size. Almost
all loops are of the order of the average intervertex distance in the network, $\sim \ln N$.
The problem of the intervertex distance statistics for these random network is reduced to
the solution of the functional equation (\ref{40}). 
It is possible to solve it only in some particular cases. 
However, all the asymptotic properties of the
distance distributions may be extracted from this equation. 
Undefined constants in the resulting asymptotic expressions (\ref{1893})--(\ref{1977}) can be found numerically, if necessary.


The general results may be summarized as follows:

\begin{enumerate} 

\item  
The average distance $\bar{d}$ between two vertices in the giant
connected component of the network depends on the network size $N$ as  
\begin{equation}
\bar{d}=\frac{\ln \left( AN\right) }{\ln z_1}\,,\;z_1=\frac{\overline{q^2}}{%
\bar{q}}-1
\,,  
\label{c10}
\end{equation}
where $A$ is some number. We assume $z_1>1$, which ensures the existence of
a giant connected component.

\item  
The mean square deviation of the intervertex distance is some finite
number $\sigma =\sqrt{\,\overline{\left( d-\bar{d}\right) ^2}}$. That is, in the large network almost all vertices in the giant
connected component are nearly equidistant from each other: the distance is
almost certainly $\bar{d}$ plus or minus a few links.

\item  
The (cumulative) intervertex distance distribution is actually a
function of $d-\bar{d}$, ${\cal Q}\left( d-\bar{d}\right) $. It is nearly $0$
as its argument is large and negative, $d<\bar{d}$, $\left| d-\bar{d}\right|
\gg 1$, and tends to $m_\infty ^2$ (the probability that two randomly chosen
vertices belong to the giant connected component) as $d>\bar{d}$, $d-\bar{d}%
\gg 1$. (Note that the narrowness of an intervertex distance distribution also was observed in other types of networks \cite{dgm02,sak02}.) 

\item  
At large negative $l=d-\bar{d}$, the asymptotics of ${\cal Q}\left(
l\right) $ is ${\cal Q}\left( l\right) \sim z_1^l$. This result is evident. 
Indeed, the average size of the $n$-th connected
component of a vertex is $\bar{m}_n\sim z_1^n$, which holds when $\bar{m%
}_n=\bar{M}_n/N\ll 1$ and these components are tree graphs.

\item  
The asymptotics of ${\cal Q}\left( l\right) $ at large positive $l$
depends on the minimal vertex degree (we assume
this degree is nonzero in any case). If $q_{\min } \equiv k=1$, we have $1-{\cal Q}%
\left( l\right) \sim lz_c^l$, where $z_c<1$ is some positive number (see the
beginning of the Section \ref{isum}). If $k=2$, the asymptotics is $1-{\cal Q}%
\left( l\right) \sim lz_c^{2l}$ with the same $z_c$. So, in these two cases
the probability that the distance between two vertices is essentially
larger than its average, decays essentially as the exponent of the
deviation. If, however, $k\ge 3$, the situation is different: $1-{\cal Q}%
\left( l\right) \sim z_1^{l/2}\exp \left( -Hz_1^{l/\left( 2-\beta \right)
}\right) $, where $\beta =\ln \left( k-1\right) /\ln z_1<1$ and $H$ is some
positive number. So, this decay is essentially more rapid than an exponential one. 
The origin of this difference is clear.  
In the first case, $k=1$, the giant connected component contains
some number of dead ends; also, when $k=2$, long chains
of vertices are present in the giant connected component. 
But, contrastingly, when $%
k\ge 3$, the giant connected component is compact.

\item  
We obtain asymptotic expressions for the size distribution $%
{\cal P}_n\left( m\right) $ for the $n$-th connected component. 
From this basic distribution, another valuable information about 
the structure of the network can be obtained. For example, 
from ${\cal P}_n\left( m\right)$, one can obtain the length 
distribution for a closed loops in the network.   

\end{enumerate}

In Sections \ref{exp} and \ref{power} our general formalism was applied to
networks with specific types of degree distribution function. In Section 
\ref{exp} we considered a two-parameter family of degree distributions. We chose $\Pi \left( q\right) \sim \zeta ^q/q$ for $q\ge 2$, $%
\zeta <1$, $\Pi \left( 1\right) =\beta <1$. A motivation for such a choice was
that the function $\phi _1\left( x\right) $ is a linear-rational one, which
allowed us to solve the main equation (\ref{40}) analytically.

In Section \ref{power} we studied the problem: what are the
statistics of intervertex distances in the networks with the 
finite first and divergent second moments of the degree distribution? We introduced a degree distribution, which behaves as $\Pi \left( q\right) \sim q^{-\gamma }\zeta ^q$%
, $2<\gamma <3$, at large degrees $q$. So, the degree 
distribution is a power law one in the limit $\zeta \rightarrow 1$. We found the leading contributions to
the size distribution for the $n$-th connected component and,
consequently, to the intervertex distance distribution in this limit. The
results may be summarized as follows:

\begin{enumerate}

\item  
The mean intervertex distance is given by: 
\begin{equation}
\bar{d}=\frac{\ln N}{(3-\gamma )\left| \ln \left( 1-\zeta \right) \right| }+2%
\frac{\ln \left| C\ln \left( 1-\zeta \right) \right| }{\left| \ln \left(
\gamma -2\right) \right| }
\, .  
\label{c20}
\end{equation}
Note that here $N\rightarrow \infty $, and $1-\zeta $ is kept small but
finite, so the first term on the right-hand part is the leading one.

\item  
The intervertex distance distribution actually depends on $%
l=d-\bar{d}$, and the form of this dependence (see Eq. (\ref{2790}))
appears to be independent of the cut-off position.

\item  
The mean square deviation of the distances (Eq. (\ref{2910})) is
again a finite number, it depends neither on the network size nor on the
cut-off.

\item  
The probability to find a pair of vertices 
separated by a distance essentially larger 
than $\bar{d}$ exponentially decays with $l=d-\bar{d}$, or, more
precisely, as $l\left( \gamma -2\right) ^{2l}$. 
The probability to find a pair 
separated by a distance essentially smaller 
than $\bar{d}$ decays faster than an exponent,
namely as $\left( \gamma -2\right) ^{l/4}\exp \left[ -2\left( \gamma
-2\right) ^{l/2}\right] $ (here $l<0$, $\left| l\right| \gg 1$).


\end{enumerate}

Now let us discuss the problem: under what conditions these results remain
true if one simultaneously tends to infinity both the size of the system and
the position of the cut-off in the degree distribution. That is, simultaneously, $N\rightarrow \infty $ and $%
\zeta \rightarrow 1$. The main question studied in this paper is: 
how does the size distribution of the $n$-th connected component changes 
with its number $n$? 
In $Z$-representation, 
at sufficiently small $n$, this evolution is described by Eq. (\ref{30}). We replace this equation with Eq. (%
\ref{45}), provided that the function $f$ is a solution of the functional
equation (\ref{40}). This can be done if, on the one hand, $n$ is large
enough, so that Eq. (\ref{30}) may be replaced with its asymptotic form,
and, on the other hand, $n$ is small enough---the size of the connected
component is still essentially smaller than the size of the network. 

Let us 
estimate how large $n$ must be to satisfy the first requirement. For this, 
let us choose
some $t_0$ close enough to $1$, so that $\phi _1\left( t_0\right) \approx
1-z_1\left( 1-t_0\right) $. This means that the second term of the
Taylor series of $\phi _1\left( t_0\right) $ near $t_0=1$, $\left(
1/2\right) \phi _1^{\prime \prime }\left( 1\right) \left( 1-t_0\right) ^2$,
is smaller than the first one. 
Since $z_1=\phi _1^{\prime }\left( 1\right)
\sim \left( 1-\zeta \right) ^{\gamma -3}$ and $\phi _1^{\prime \prime
}\left( 1\right) \sim \left( 1-\zeta \right) ^{\gamma -4}$, this is
satisfied if at least $1-t_0\lesssim 1-\zeta $. The condition for the
replacement of the evolution equation (\ref{30}) with its asymptotical form
means that the functions $F_n\left( x\right) $ and $F_{n+1}\left( x\right) $
can be reduced to each other by the rescaling of the independent variable, $%
F_{n+1}\left( x\right) =F_n\left( z_1x\right) $. This is true, if after $n$
iterations of the interval of linearity of the function $\phi _1\left( x\right) $, $%
\left( t_0,1\right) $, $t_0\sim 1-\zeta $, the resulting interval $\left(
t_n,1\right) $ nearly coincides with its limit at $n\rightarrow \infty $, $%
\left( 0,1\right) $. In other words, we must require $t_n\ll 1$. Outside the
interval of linearity we can write: 
\[
1-t_{n+1}=1-\phi _1\left( t_n\right) \rightarrow \left( 1-t_n\right)
^{\gamma -2}\,. 
\]
Consequently $t_n=1-\left( 1-t_0\right) ^{\left( \gamma -2\right) ^n}\cong \left(
\gamma -2\right) ^n\left| \ln \left( 1-t_0\right) \right| \sim \left( \gamma
-2\right) ^n\left| \ln \left( 1-\zeta \right) \right| \ll 1$. Hence  
\begin{equation}
n\gg \ln \left| \ln \left( 1-\zeta \right) \right| 
\,.  
\label{c30}
\end{equation} 

On the other hand, the average size of the $n$-th connected component, $\bar{M%
}_n\sim z_1^n\sim \left( 1-\zeta \right) ^{-\left( 3-\gamma \right) n}$, must
be essentially smaller than the network size $N$. This imposes the limitation: 
\begin{equation}
n\ll \frac{\ln N}{\left| \ln \left( 1-\zeta \right) \right| }
\,.  
\label{c40}
\end{equation}
Combining inequalities (\ref{c30}) and (\ref{c40}), we get $\ln N\gg \left|
\ln \left( 1-\zeta \right) \right| \ln \left| \ln \left( 1-\zeta \right)
\right| $, or the restriction to the range of values of the cut-off parameter, where our approach is valid: 
\begin{equation}
\left| \ln \left( 1-\zeta \right) \right| \ll \frac{\ln N}{\ln \ln N}\,.
\label{c50}
\end{equation} 

Let us introduce an $N$-dependent cut-off, which growth with $N$, being on
the boundary of applicability of our approach, i.e. 
$|\ln [1-\zeta(N)]| \sim \ln N/\ln \ln N$. 
In this event, instead of Eq. (\ref{c20}), we obtain  
\begin{equation}
\bar{d}\sim \ln \ln N
\,.  
\label{c60}
\end{equation} 
Recall that here $2 < \gamma < 3$. Also, recall that the position (degree) of the cut-off, $q_0$, and the parameter $\zeta$ are related in the following way: $\zeta= e^{-1/q_0}$. So, the boundary of applicability of the approach is $\ln q_0 \sim \ln N/\ln\ln N$.) 

For finite networks, a fat-tailed degree distribution function, 
necessarily has a cut-off, whose dependence on the network size is 
determined by the
details of the construction procedure. If this dependence satisfies the
condition (\ref{c50}), the results of Section \ref{power} are
applicable. That is, as the network size increases, 
the intervertex distance distribution eventually assumes $N$-independent
shape, but centered at $\bar{d}\left( N\right) $, where $\ln \ln N\ll 
\bar{d}\left( N\right) \ll \ln N$. 
When the position (degree) of the cut-off grows with $N$ faster than that on the boundary of applicability of the approach, then $\bar{d}$ increases with $N$ even more slowly than in Eq. (\ref{c60}). 

Let us obtain an estimate of the asymptotic dependence $\bar{d}(N)$, as $N \to \infty$, 
in the situation when $2<\gamma<3$ and the cutoff degree grows with $N$ as a power law, $q_0 \propto N^\epsilon$. Here, $0<\epsilon<1$. 
(Again we assume that the giant connected component coincides with the entire network.)  
This behaviour corresponds to $1-\zeta(N) \sim N^{-\epsilon}$. From Eq.~(\ref{i10}), we see that 
$z_1(N) \sim N^{\epsilon(3-\gamma)}$. 
The size of the vertex component approaches a value of a finite fraction of $N$ in $l_1$ steps: 
${\rm const}_1 z_1^{l_1} \sim {\rm const}_2 N^{\epsilon(3-\gamma) l_1} 
\sim {\rm const}_3 N$. So, the ``linear'' stage of the vertex component evolution is completed in a finite number of steps, 
$l_1 \sim 1/[\epsilon(3-\gamma)]$. 

After this, a point $t_n$ in the mapping $t_n \to t_{n+1}$ [see Eq. (\ref{140})] appears in the region $1-t_{0'} \sim 1- \zeta \ll 1$. Speaking more precisely, the last step of the ``linear'' stage cannot immediately make $1-t_{0'} \sim 1$, since it leads, at the worst, to the multiplication of the value of $1-t_{0'}$ by $z_1$ which gives 
$N^{\epsilon(3-\gamma)}N^{-\epsilon} \sim N^{\epsilon(\gamma-2)} \ll 1$. 
So, there is a space for the second stage of the vertex component evolution, 
which is described by the mapping $1-t_{n+1}=(1-t_n)^{\gamma-2}$. 
Consequently, $1-t_n$ approaches values $1-t \sim 1$ in 
$l_2 \cong \ln\!|\!\ln(1-t_{0'})|\, /|\ln(\gamma-2)| \sim \ln\ln N/|\ln(\gamma-2)|$ steps. 
$\bar{d} \approx l_1+l_2$, so that the mean intervertex distance of networks with degree distribution exponent $2 < \gamma < 3$ and a power-law $q_0(N)$ is   
\begin{equation}
\bar{d}(N) \sim \frac{\ln\ln N}{|\ln(\gamma-2)|\phantom{\!\!\!\!\!\!\!I^I}}
\, .  
\label{c101}
\end{equation} 

If the $\gamma$ exponent of the degree distribution is equal to $3$ and $q_0 \propto N^\epsilon$, then $ z_1 \propto \ln N$, see Eq. (\ref{i10}).  
The size of the vertex component approaches a value of a finite fraction of $N$ in $l_1$ steps: 
${\rm const}_1 z_1^{l_1} \sim {\rm const}_2 (\ln N)^{l_1} 
\sim {\rm const}_3 N$. So, the ``linear'' stage of the vertex component evolution is completed in $l_1 \sim \ln/\ln \ln N$ steps. 
After this, a point $t_n$ in the mapping $t_n \to t_{n+1}$  [see Eq. (\ref{140})] appears in the region $1-t_{0'} \sim 1- \zeta \sim N^{-\epsilon} \ll 1$. 
The $Z$ transform of the function with the asymptote $k^{-3}$ at large $k$ is $\phi(z) \cong \frac{1}{2}(1-z)^2 |\ln (1-z)|$ near $z=1$, so that 
$\phi_1(z) \cong 1-(1-z)|\ln (1-z)|$. So, in the region $1-\zeta(N)<1-t_n \ll 1$, the mapping is 
\begin{equation}
1-t_{n+1} = (1-t_n) |\ln (1-t_n)|
\, .  
\label{c102}
\end{equation} 
It takes $l_2 \cong |\ln (1-t_{0'})|/\ln\!|\ln (1-t_{0'})| 
\sim \epsilon \ln N/\ln\ln N$ steps to approach $1-t_n \sim 1$. 
Consequently, the mean intervertex distance of networks with degree distribution exponent $\gamma=3$ and a power-law $q_0(N) \propto N^\epsilon$ behaves as 
\begin{equation}
\bar{d}(N) \sim (1+\epsilon)\frac{\ln N}{\ln\ln N}
\, .  
\label{c103}
\end{equation} 

The estimates (\ref{c101}) and (\ref{c103}) support those in Ref. \cite{ch02}. 
Note that the resulting asymptotics of the mean intervertex distance (\ref{c101}) and (\ref{c103}) are independent of the mean degree. 
This feature may be checked by more detailed calculations. Moreover, in the case $2<\gamma<3$, the expression (\ref{c101}) for the mean intervertex distance does not contain exponent $\epsilon$. 

In summary, we have developed a consistent formalism for the calculation of statistical characteristics of uncorrelated random networks with an arbitrary degree distribution. 
This formalism accounts for the complex structure of such networks. 
We mainly focused on the intervertex distance distributions, but many other distributions may be studied in a similar way.    

S.N.D. thanks PRAXIS XXI (Portugal) for a research grant PRAXIS XXI/BCC/16418/98. S.N.D. and J.F.F.M.  
were partially supported by the project POCTI/99/FIS/33141. 
A.N.S. acknowledges the NATO program OUTREACH for support. 
We also thank V.V.~Bryksin, A.V.~Goltsev, and A.~Krzywicki for useful discussions. 

\appendix 

\section{Characteristics of the distance distribution function}
\label{q}

Taking into account the properties of the functions $g\left( y\right) $ and $%
p\left( x\right) $, one can write: 
\begin{equation}
m_\infty ^2-{\cal Q}\left( l\right) =m_\infty ^2\int_0^\infty dx\,\tilde{p}%
\left( x\right) \tilde{g}\left( z_1^lx\right) 
\,,\,  
\label{270}
\end{equation}
where $\tilde{g}$ is defined in Eq. (\ref{182}) and $\tilde{p}$ is its
inverse Laplace transform: 
\begin{equation}
\tilde{g}\left( y\right) =\int_0^\infty dx\,\tilde{p}\left( x\right)
e^{-xy}\,,\;\tilde{p}\left( x\right) =\int_{-i\infty }^{+i\infty }\frac{dy}{%
2\pi i}\tilde{g}\left( y\right) e^{xy}
\,.  
\label{280}
\end{equation}
We consider the asymptotics of the distance distribution function ${\cal Q}\left( l\right) $ at large positive $l$,
i.e. at small $\nu _l=z_1^{-l}$. Let us choose some $x_0$ satisfying the conditions: $\nu _l\ll x_0\ll 1$. Within the interval $\left( 0,x_0\right) $, 
one can replace $\tilde{p}\left( x\right) $ with $Bx^{\alpha -1}$ in the
integral in Eq. (\ref{270}), and one can replace $\tilde{g}\left(
y/\nu _l\right) $ with $B\left( y/\nu _l\right) ^{-\alpha }$ within the
interval $\left( x_0,\infty \right) $. Thus we have: 
\begin{equation}
m_\infty ^2-{\cal Q}\left( l\right) \cong m_\infty ^2\frac B{\Gamma \left(
\alpha \right) }\int_0^{x_0}dx\,x^{\alpha -1}\tilde{g}\left( x/\nu _l\right)
+m_\infty ^2B\nu _l^\alpha \int_{x_0}^\infty dx\,x^{-\alpha }\tilde{p}\left(
x\right) 
\,.  
\label{290}
\end{equation}
Let us replace the integration variable $x$ in the first integral with $y=x/\nu
_l$. Then, in the first integral we represent the integrand as $d\left( \ln
y\right) \,y^\alpha \tilde{g}\left( y\right) $ and integrate by parts. In
the second integral we also write the integrand as $d\left( \ln x\right)
\,x^{1-\alpha }\tilde{p}\left( x\right) $ and integrate by parts. As a
result, we have: 
\begin{eqnarray}
& & m_\infty ^2-{\cal Q}\left( l\right) \cong  
\nonumber 
\\[5pt]
m_\infty ^2B\nu _l^\alpha & & 
\left\{  \frac 1{\Gamma \left( \alpha \right) }\tilde{g}\left( x_0/\nu
_l\right) \left( \frac{x_0}{\nu _l}\right) ^\alpha \ln \left( \frac{x_0}{\nu
_l}\right) -\frac 1{\Gamma \left( \alpha \right) }\int_0^{x_0/\nu
_l}dy\,y^{\alpha -1}\ln y\left[ y\tilde{g}^{\prime }\left( y\right) +\alpha 
\tilde{g}\left( y\right) \right] -\right.  
\nonumber 
\\[5pt]
& & \left. \tilde{p}\left( x_0\right) x^{1-\alpha }\ln x_0-\int_{x_0}^\infty
dx\,x^{-\alpha }\ln x\left[ x\tilde{p}^{\prime }\left( x\right) -\left(
\alpha -1\right) \tilde{p}\left( x\right) \right]\right\} 
\,.  
\label{300}
\end{eqnarray}
One can replace $\tilde{g}\left( x_0/\nu _l\right) $ in the first integrated
term with its asymptotics $B\left( x_0/\nu _l\right) ^{-\alpha }$, because $%
x_0/\nu _l\gg 1$, and $\tilde{p}\left( x_0\right) $ with $Bx_0^{\alpha
-1}/\Gamma \left( \alpha \right) $ in the second integrated term, because $%
x_0\ll 1$. Also, the upper limit in the first integral may be replaced with $%
\infty $, because this integral is convergent, $y\tilde{g}^{\prime }\left(
y\right) +\alpha \tilde{g}\left( y\right) ={\cal O}\left( 1/y\right) $ as $%
y\rightarrow \infty $. Similarly, the lower limit in the second integral may
be replaced with $0$, because $x\tilde{p}^{\prime }\left( x\right) -\left(
\alpha -1\right) \tilde{p}\left( x\right) ={\cal O}\left( x\right) $ as $%
x\rightarrow 0$. So,  
\begin{eqnarray}
m_\infty ^2-{\cal Q}\left( l\right) & \cong & \frac{m_\infty ^2B\nu _l^\alpha 
}{\Gamma \left( \alpha \right) }\left\{ B\ln \left( \frac 1{\nu _l}\right)
-\int_0^\infty dy\,y^{\alpha -1}\ln y\left[ y\tilde{g}^{\prime }\left(
y\right) +\alpha \tilde{g}\left( y\right) \right] - \right. 
\nonumber 
\\[5pt]
& & \left. \Gamma \left( \alpha
\right) \int_0^\infty dx\,x^{-\alpha }\ln x\left[ x\tilde{p}^{\prime }\left(
x\right) -\left( \alpha -1\right) \tilde{p}\left( x\right) \right] \right\}
\,.  
\label{310}
\end{eqnarray}
Each of the two integrals in the braces may be expressed in terms of the
other one. For example, let us take the first integral, and substitute into it  
$\tilde{g}(y)$, which is the Laplace transform of $\tilde{p}(x)$, Eq. (\ref{280}). Note that  
\[
y\tilde{g}^{\prime }\left( y\right) =-\int_0^\infty dx\,\left[ x\tilde{p}%
^{\prime }\left( x\right) +\tilde{p}\left( x\right) \right] e^{-xy}\,. 
\]
Then, changing the order of integrations, we have  
\begin{eqnarray}
\int_0^\infty dy\,y^{\alpha -1}\ln y\left[ y\tilde{g}^{\prime }\left(
y\right) +\alpha \tilde{g}\left( y\right) \right] &=&-\int_0^\infty dx\left[
x\tilde{p}^{\prime }\left( x\right) -\left( \alpha -1\right) \tilde{p}\left(
x\right) \right] \int_0^\infty dy\,y^{\alpha -1}e^{-xy}\ln y=  
\nonumber \\[5pt]
& &\Gamma \left( \alpha \right) \int_0^\infty dx\,x^{-\alpha }\left[ \ln
x-\psi \left( \alpha \right) \right] \left[ x\tilde{p}^{\prime }\left(
x\right) -\left( \alpha -1\right) \tilde{p}\left( x\right) \right] 
\,,
\label{320}
\end{eqnarray}
where $\psi(\alpha)=[\ln\Gamma(\alpha)]'$.
Quite analogously,  
\begin{equation}
\int_0^\infty dy\,y^{\alpha -1}\left[ y\tilde{g}^{\prime }\left( y\right)
+\alpha \tilde{g}\left( y\right) \right] =-\Gamma \left( \alpha \right)
\int_0^\infty dx\,x^{-\alpha }\left[ x\tilde{p}^{\prime }\left( x\right)
-\left( \alpha -1\right) \tilde{p}\left( x\right) \right] 
\,.  
\label{330}
\end{equation}
Then from Eqs. (\ref{320}) and (\ref{330}) one can easily obtain  
\begin{equation}
\int_0^\infty dx\,x^{-\alpha }\ln x\left[ x\tilde{p}^{\prime }\left(
x\right) -\left( \alpha -1\right) \tilde{p}\left( x\right) \right]
=\int_0^\infty dy\,y^{\alpha -1}\left[ \ln y-\psi \left( \alpha \right)
\right] \left[ y\tilde{g}^{\prime }\left( y\right) +\alpha \tilde{g}\left(
y\right) \right] 
\,.  
\label{340}
\end{equation}
Substituting either Eq. (\ref{330}) or Eq. (\ref{340}) into Eq. (\ref{310})
and taking into account that $\nu _l^\alpha =z_1^{-\alpha l}=z_c^l$, we
arrive at expressions represented by Eqs. (\ref{193})--(\ref{194}).

Now let us calculate the first two moments of $l$. We have: 
\begin{equation}
\bar{l}=-\int_0^\infty dx\,\tilde{p}\left( x\right) \int_{-\infty }^{+\infty
}dl\,l\frac \partial {\partial l}\tilde{g}\left( z_1^lx\right) 
\,.
\label{350}
\end{equation}
Changing the integration variable $l$ by $y=z_1^lx$, $l=\left( \ln y-\ln
x\right) /\ln z_1$, we obtain  
\begin{equation}
\bar{l}=-\frac 1{\ln z_1}\left[ \int_0^\infty dy\,\tilde{g}^{\prime }\left(
y\right) \ln y+\int_0^\infty dx\,\tilde{p}\left( x\right) \ln x\right] 
\,.
\label{360}
\end{equation}
Again, the integrals in the Eq. (\ref{360}) can be mutually expressed. Substituting $\tilde{g}$ 
from Eq. (\ref
{280}) and changing the order of integrations, we get  
\begin{equation}
\int_0^\infty dy\,\tilde{g}^{\prime }\left( y\right) \ln y=-\int_0^\infty
dx\,\tilde{p}\left( x\right) x\int_0^\infty dy\,e^{-xy}\ln y=\int_0^\infty
dx\,\tilde{p}\left( x\right) \ln x+\gamma_e
\,,  
\label{370}
\end{equation}
where $\gamma_e $ is the Euler constant. So, we have obtained the
expression (\ref{196}) for $\bar{l}$. Analogously we have 
\begin{equation}
\overline{l^2}=-\frac 1{\ln ^2z_1}
\left[ \int_0^\infty\! dy\,\tilde{g}^{\prime
}\left( y\right) \ln ^2y-\int_0^\infty\! dx\,\tilde{p}\left( x\right) \ln
^2x-2\!\int_0^\infty\! dy\,\tilde{g}^{\prime }\left( y\right) \ln y\int_0^\infty
\!dx\,\tilde{p}\left( x\right) \ln x\right] 
.  
\label{380}
\end{equation}
Quite similarly to Eq. (\ref{370}) one can obtain  
\begin{equation}
\int_0^\infty dy\,\tilde{g}^{\prime }\left( y\right) \ln ^2y=-\int_0^\infty
dx\,\tilde{p}\left( x\right) \ln ^2x-2\gamma \int_0^\infty dx\,\tilde{p}%
\left( x\right) x-\gamma ^2-\frac{\pi ^2}6
\,.  
\label{390}
\end{equation}
Then, using Eqs. (\ref{370}) and (\ref{390}), one can rewrite Eq. (\ref{380}%
) as  
\begin{eqnarray}
& & \overline{l^2} = 
\nonumber 
\\[5pt] 
& & \frac 1{\ln ^2z_1}\left\{ 2\!\int_0^\infty\! dx\,\tilde{p}%
\left( x\right) \ln ^2x+4\gamma\! \int_0^\infty \!dx\,\tilde{p}\left( x\right)
\ln x+2\left[ \int_0^\infty\! dx\,\tilde{p}\left( x\right) \ln x\right]
^2\!+\gamma ^2+\frac{\pi ^2}6\right\} =  
\nonumber 
\\[5pt]
& &-\frac 1{\ln ^2z_1}\left\{ \int_0^\infty\! dy\,\tilde{g}^{\prime }\left(
y\right) \ln ^2y+4\gamma \int_0^\infty\! dy\,\tilde{g}^{\prime }\left(
y\right) \ln y-2\left[ \int_0^\infty\! dy\,\tilde{g}^{\prime }\left( y\right)
\ln y\right] ^2\!-\gamma ^2+\frac{\pi ^2}6\right\}  
\, . 
\label{400}
\end{eqnarray}
From Eqs. (\ref{360}) and (\ref{400}), we immediately obtain Eq. (\ref{197}).

\section{Solution of the main equation in the case of  
a power-law degree distribution with an exponential cut-off}
\label{saw}

Differentiating Eq. (\ref{2660}) twice, setting $y=0$, and taking into
account the initial conditions for $f\left( y\right) $, one can get the
expression for $f^{\prime \prime }\left( 0\right) $: 
\begin{equation}
f^{\prime \prime }\left( 0\right) =\frac{\left( 3-\gamma \right) \zeta }{%
\left( z_1-1\right) \left( 1-\zeta \right) }\cong \frac{3-\gamma }{\gamma
-2}\left( 1-\zeta \right) ^{2-\gamma } 
\,.  
\label{410}
\end{equation}
This means one can use the linear approximation for $f(y)$, $f\left( y\right)
\approx 1-y$, as long as $\left| y\right| \ll \left( 1-\zeta \right)
^{\gamma -2}$.
 The next step is to make one iteration in Eq. (\ref{2660}) 
substituting $f\left( y\right) =1-y$ into its right-hand side. The resulting expression
for $f\left( y\right) $,  
\begin{eqnarray}
f\left( y\right) &=&\left[ 1-\left( 1-\zeta \right) ^{\gamma -2}\right]
^{-1}\left[ 1-\left( 1-\zeta +\frac \zeta {z_1}y\right) \right] \cong  
\nonumber \\[5pt]
& & 1-\left( 1-\zeta \right) ^{\gamma -2}\left[ \left( 1+\frac y{%
\left( \gamma -2\right) \left( 1-\zeta \right) ^{\gamma -2}}\right) ^{\gamma
-2}-1\right] 
\,,  
\label{420}
\end{eqnarray}
is valid if $\left| y\right| \ll \left( 1-\zeta \right) ^{\gamma -2}z_1\sim
\left( 1-\zeta \right) ^{2\gamma -5}$. In particular, if $\left( 1-\zeta
\right) ^{\gamma -2}\ll \left| y\right| \ll \left( 1-\zeta \right) ^{2\gamma
-5}$, Eq. (\ref{420}) may be reduced to  
\begin{equation}
f\left( y\right) \cong 1-\left[ \frac y{\left( \gamma -2\right) \left(
1-\zeta \right) ^{\gamma -3}}\right] ^{\gamma -2}
\,.  
\label{430}
\end{equation}

On the other hand, in this region we have $\left| 1-f\left( y\right)
\right| \gg \left( 1-\zeta \right) ^{\gamma -2}$ and $\left| 1-f\left(
x/z_1\right) \right| \ll 1-\zeta $. 
Therefore, if $\left| y\right| \gg
\left( 1-\zeta \right) ^{\gamma -2}$, one can neglect the term $\left(
1-\zeta \right) ^{\gamma -2}f\left( y\right) $ on the left-hand side of Eq. (\ref{2660}%
), and the term $\left( 1-\zeta \right) f\left( y/z_1\right) $ on the right-hand side,
i.e. simply to set $\zeta =1$ in this equation. As a result, we obtain Eq. (%
\ref{2680}), whose solution is given by Eq. (\ref{2690}) with the constant $%
A $ to be determined. This can be done, if we take into account that in the
interval $\left( 1-\zeta \right) ^{\gamma -2}\ll \left| y\right| \ll \left(
1-\zeta \right) ^{2\gamma -5}$, both Eq. (\ref{430}) and Eq. (\ref{2690}) are
valid. 
Let us choose some $y_0$ within this interval and represent there
the exponent in Eq. (\ref{2660}) as $Ay^{-\vartheta }=Ay_0^{-\vartheta }\left(
1-\vartheta \ln \left( y/y_0\right) \right) $. Then one can write  
\begin{equation}
f\left( y\right) =1-\left( \frac y{y_0}\right) ^{\vartheta Ay_0^{-\vartheta }}\exp
\left( -Ay_0^{-\vartheta }\right) \,.  \label{440}
\end{equation}
Comparing Eqs. (\ref{430}) and (\ref{440}), we obtain: 
\begin{equation}
x_0=e(\gamma-2)\left( 1-\zeta \right) ^{\left( \gamma -3\right) \left[ 1+1/\ln \left(
\gamma -2\right) \right] }\,,\;A=\frac 1{e\vartheta }
\,.  
\label{450}
\end{equation}
Here the expression (\ref{2690}) for $\vartheta $ was used.

Thus we obtain Eq. (\ref{2710}) for $g\left( y\right) =\phi \left[ f\left(
y\right) \right] $, which is valid if $\left| y\right| \gg \left( 1-\zeta
\right) ^{\gamma -2}$. Now we must calculate its inverse Laplace
transform. That is, we have to calculate the integral  
\begin{equation}
\int_{-i\infty +\delta }^{+i\infty +\delta }\frac{dy}{2\pi i}e^{xy}\left[
1-\exp \left( -\frac{y^{-\vartheta }}{e\vartheta }\right) \right] =-\int_c\frac{dy%
}{2\pi i}\exp \left( xy-\frac{y^{-\vartheta }}{e\vartheta }\right) 
\,,  
\label{460}
\end{equation}
and the other one which differs only by the additional multiple $\gamma -1$
before the second term in the exponential. The integration contour $c$ goes from $%
-\infty $ to $0$ along the lower shore of the $\left( -\infty ,0\right) $
cut and comes back along the upper shore. This equality holds as $x>0$. Note 
that the term in the braces on the left-hand side of Eq. (\ref{460}) becomes
zero as $y\rightarrow +\infty $, which means that the integral has no $%
\delta $-functional singularity at $x=0$. We assume here $\left| x\right|
\ll \left( 1-\zeta \right) ^{2-\gamma }$ and will make use of the smallness
of $\vartheta $. The actual region of integration is $\left| y\right| \lesssim
1/x$. Here we effectively write: 
\[
\frac{y^{-\vartheta }}{e\vartheta }=\frac{x^\vartheta }{e\vartheta }\left( xy\right)
^{-\vartheta }\approx \frac{x^\vartheta }{e\vartheta }\left[ 1-\vartheta \ln \left(
xy\right) \right] 
\,. 
\]
Indeed, because this expression is in an exponential, the criterion is the
smallness of the neglected terms compared to $1$. We can estimate them as $%
\vartheta x^\vartheta \ln ^2\left( xy\right) \sim \vartheta x^\vartheta <\vartheta \left(
1-\zeta \right) ^{\vartheta \left( 2-\gamma \right) }\sim \vartheta \ll 1$. Then
the integral in Eq. (\ref{460}), after the replacement of the integration
variable, $z=xy$, turns into  
\begin{equation}
\frac{\exp \left( -x^\vartheta /e\vartheta \right) }x\int_c\frac{dz}{2\pi i}%
z^{x^\vartheta /e}e^x=\left[ x\Gamma \left( -\frac{x^\vartheta }e\right) \right]
^{-1}\exp \left( -\frac{x^\vartheta }{e\vartheta }\right) 
\,.  
\label{470}
\end{equation}
The function on the right-hand side of Eq. (\ref{470}) is essentially nonzero if $%
x^\vartheta \lesssim \vartheta $, so that one can replace there $\Gamma \left(
-x^\vartheta /e\right) \rightarrow -ex^{-\vartheta }$\vspace{10pt}. 

\noindent
$^{*}$ Electronic address: sdorogov@fc.up.pt \newline
$^{\dagger }$ Electronic address: jfmendes@fc.up.pt \newline
$^{\ddag }$ Electronic address: samukhin@fc.up.pt

\end{document}